\documentclass[a4paper]{aa}

\usepackage{graphics,epsfig,txfonts}
\usepackage[section]{placeins}
\usepackage{natbib}
\bibpunct{(}{)}{;}{a}{}{,}
\usepackage[colorlinks=true,linkcolor=black,anchorcolor=black,citecolor=black,filecolor=black,menucolor=black,runcolor=black,urlcolor=black]{hyperref}

\newcommand{\ltsima}{$\buildrel < \over \sim$}
\newcommand{\lsim}{\lower.5ex\hbox{\ltsima}}
\newcommand{\gtsima}{$\buildrel > \over \sim$}
\newcommand{\gsim}{\lower.5ex\hbox{\gtsima}}
\begin{document}
 
\title{Thermodynamic perturbations in the X-ray halo of 33 clusters of galaxies observed with \emph{Chandra} ACIS
\thanks{Tables 2-67 are only available in electronic form at the CDS via anonymous ftp to cdsarc.u-strasbg.fr (130.79.128.5) or via http://cdsweb.u-strasbg.fr/cgi-bin/qcat?J/A+A/}}

\author{      F.~Hofmann\inst{1}
     \and     J.S.~Sanders\inst{1}
     \and     K.~Nandra\inst{1}
     \and     N.~Clerc\inst{1}
     \and     M.~Gaspari\inst{2,3}
       }

\titlerunning{Thermodynamic perturbations in galaxy clusters}
\authorrunning{Hofmann et al.}

\institute{Max-Planck-Institut f\"ur extraterrestrische Physik, Giessenbachstra{\ss}e, 85748 Garching, Germany
           \and Department of Astrophysical Sciences, Princeton University, Princeton, NJ 08544, USA; Einstein and Spitzer Fellow
           \and Max-Planck-Institut f\"ur Astrophysik, Karl-Schwarzschildstra{\ss}e, 85748 Garching, Germany
          }

%\date{Received dd month 2015 / Accepted dd month 2015}

\abstract{In high-resolution X-ray observations of the hot plasma in clusters of galaxies significant structures caused by AGN feedback, mergers, and turbulence can be detected. Many clusters have been observed by \emph{Chandra} in great depth and at high resolution.}
	 {Using archival data taken with the \emph{Chandra} ACIS instrument the aim was to study thermodynamic perturbations of the X-ray emitting plasma and to apply this to better understand the thermodynamic and dynamic state of the intra cluster medium (ICM).}
 	 {We analysed deep observations for a sample of 33 clusters with more than 100\,ks of \emph{Chandra} exposure each at distances between redshift 0.025 and 0.45. The combined exposure of the sample is 8\,Ms. Fitting emission models to different regions of the extended X-ray emission we searched for perturbations in density, temperature, pressure, and entropy of the hot plasma.}
	 {For individual clusters we mapped the thermodynamic properties of the ICM and measured their spread in circular concentric annuli. Comparing the spread of different gas quantities to high-resolution 3D hydrodynamic simulations, we constrain the average Mach number regime of the sample to $\rm{Mach_{1D}\approx0.16\pm0.07}$. In addition we found a tight correlation between metallicity, temperature and redshift with an average metallicity of $\rm{Z\approx 0.3\pm0.1~Z_\odot}$.}
	 {This study provides detailed perturbation measurements for a large sample of clusters which can be used to study turbulence and make predictions for future X-ray observatories like eROSITA, Astro-H, and Athena.}

\maketitle

\section{Introduction}
\label{sec:introduction}

In the current picture of the evolution of the universe, clusters of galaxies have formed in the deep potential wells created by clumping of dark matter (DM) around remnant density fluctuations after the big bang. The majority of the mass in clusters is made of DM, which is only observed indirectly through its gravitational effects. The second component of clusters is baryonic matter consisting mainly of very thin hot plasma (the intra cluster medium, ICM), heated by the gravitational accretion into the potential wells, and emitting strongest in X-ray wavelength due to its high temperatures. The smallest fraction of the mass is in the stellar content of the clustered galaxies, which is observable in visible light. The behaviour of DM in the cluster potential is believed to be well understood from cosmological simulations \citep[e.g.][]{2005Natur.435..629S} and observations of gravitational lensing effects on visible light \citep[][]{1995ApJ...438...49B, 1995ApJ...449..460K, 2002MNRAS.334L..11A, 2006ApJ...652..937B, 2008A&A...482..451Z, 2013ApJ...767..116M}. The member galaxies of a cluster are well described as collisionless particles moving in the cluster potential by measuring the line of sight velocity dispersion in the optical \citep[see e.g.][]{2011A&A...526A.105Z, 2013ApJ...767...15R}. The complex dynamic and thermodynamic processes in the hot ICM can be studied with X-ray observations. Other phases of the ICM have been studied at different wavelength with UV and $\rm{H{\alpha}}$ \citep[e.g.][]{2011ApJ...734...95M} or radio observations \citep[e.g.][]{2001A&A...378..777D,2004ApJ...605..695G}.

\citet{2004A&A...426..387S} first related fluctuations in the projected pressure maps of the hot, X-ray emitting, ICM of the Coma cluster to turbulence. Turbulence has been discussed as a significant heating mechanism \citep[see][]{2005ApJ...622..205D,2010ApJ...713.1332R,2012MNRAS.424..190G}, which is important to understand the heating and cooling balance \citep[see cooling flow problem,][]{1994ARA&A..32..277F} in clusters and estimate the amount of non-thermal pressure support \citep[see e.g. simulations by][]{2014ApJ...792...25N}. \citet{2014Natur.515...85Z} recently studied turbulence in the Perseus and Virgo cluster by analysing fluctuations in the surface brightness of the cluster emission. Asymmetries and fluctuations within thermodynamic properties of the hot plasma can be used to estimate the amount of turbulence \citep[e.g.][]{2013A&A...559A..78G,2014A&A...569A..67G}. This has been studied in the PKS\,0745-191 galaxy cluster by \citet{2014MNRAS.444.1497S}. We applied a similar technique to the current sample of 33 clusters and compared our results to the findings of cluster simulations \citep[e.g.][]{2009A&A...504...33V,2009ApJ...705.1129L}.

The amount of turbulence in the hot ICM is hard to directly measure. Simulations of galaxy clusters predict turbulent motions of several hundreds of km/s \citep[e.g.][]{2005MNRAS.364..753D,2014ApJ...792...25N,2014A&A...569A..67G}. \citet{2011MNRAS.410.1797S}, \citet{2013MNRAS.429.2727S}, and \citet{2015A&A...575A..38P} were able to obtain upper limits on the velocity broadening of spectra for a large sample of clusters. However current X-ray instruments do not provide the spectral resolution needed to detect significant broadening due to turbulence in all but a few possible cases.

The basis for this study was the X-ray all-sky survey of the \emph{ROSAT} mission \citep[1990 to 1999, see][]{1982AdSpR...2..241T}. The clusters identified in this survey \citep[e.g.][]{2000ApJS..129..435B,2004A&A...425..367B} were followed up with the current generation X-ray telescopes \emph{Chandra}, \emph{XMM-Newton}, and \emph{Suzaku}. For the substructure study we used observations of \emph{ROSAT}-clusters with the X-ray observatory on board the \emph{Chandra} satellite which delivers the best spatial and very good spectral resolution. Since its launch in 1999 \emph{Chandra} has frequently been used to study clusters of galaxies as individual systems and as cosmological probes \citep[e.g.][]{2004MNRAS.353..457A,2008MNRAS.383..879A,2009ApJ...692.1060V, 2009ApJ...692.1033V}. Its high resolution showed an unexpected complexity of the ICM structure in many cases \citep[e.g.][]{2000MNRAS.318L..65F, 2000ApJ...534L.135M, 2000ApJ...541..542M, 2002ApJ...567L..27M, 2005MNRAS.356.1022S, 2006MNRAS.366..417F, 2007ApJ...665.1057F}. Based on the large archive of deep cluster observations we analysed a sample of clusters and mapped their thermodynamic properties. We specifically investigated perturbations in the thermodynamic parameters of the ICM, which according to recent high-resolution simulations can be used to trace turbulence in the ICM via the normalisation of the ICM power spectrum \citep[e.g. in density,][]{2014A&A...569A..67G}. By measuring the slope of the power spectrum we can constrain the main transport processes in the hot ICM, such as thermal conductivity 
\citep[][]{2013A&A...559A..78G}.

The paper is structured as follows: Sect. 2 describes the sample selection and data reduction, Sect. 3 describes the analysis of perturbations, in Sect. 4 the main results are presented, Sect. 5 contains results for individual systems, in Sect. 6 the findings are discussed, and Sect. 7 contains the conclusions.

For all our analysis we used a standard $\rm{\Lambda CDM}$ cosmology with $\rm{H_0=71~km~s^{-1}~Mpc^{-1}}$, $\rm{\Omega_M=0.27}$ and $\rm{\Omega_{\Lambda}=0.73}$ and relative solar abundances as given by \citet{1989GeCoA..53..197A}.

\section{Observations and data reduction}
\label{sec:observations}
%--------------------------
\begin{table*}[]
\caption[]{Sample properties.}
\begin{center}
\begin{tabular}{llrrrrrr}
\hline\hline\noalign{\smallskip}
  \multicolumn{1}{l}{Cluster\tablefootmark{a}} &
  \multicolumn{1}{l}{Abbrev.\tablefootmark{a}} &
  \multicolumn{1}{l}{$\rm{n_H}$} &
  \multicolumn{1}{l}{<T map>} &
  \multicolumn{1}{l}{$\rm{r_{FOV}}$\tablefootmark{b}} &
  \multicolumn{1}{l}{$\rm{r_{500}}$\tablefootmark{c}} & 
  \multicolumn{1}{l}{$\rm{M_{500}}$\tablefootmark{c}} & 
  \multicolumn{1}{l}{$\rm{r_{FOV}/r_{500}}$} \\
  \multicolumn{1}{l}{} &
  \multicolumn{1}{l}{} &
  \multicolumn{1}{l}{$\rm{[10^{22}~cm^{-2}]}$} &
  \multicolumn{1}{l}{[keV]} &
  \multicolumn{1}{l}{[Mpc]} &
   \multicolumn{1}{l}{[Mpc]} &
   \multicolumn{1}{l}{$\rm{[10^{14}~M_\odot]}$} &
   \multicolumn{1}{l}{} \\
\noalign{\smallskip}\hline\noalign{\smallskip}
      
RX J1347-114 & rxj1347 & 0.046 & 14.7$\pm$0.125 & 0.48 & 1.57$\pm$0.05  & 17.7$\pm$1.8 & 0.31 \\   
1E 0657-56 & 1e0657 & 0.049 & 12.6$\pm$0.112 & 0.98 & 1.58$\pm$0.06  & 15.3$\pm$1.9 & 0.62 \\   
A 2390 & a2390 & 0.062 & 10.8$\pm$0.104 & 0.55 & 1.51$\pm$0.07  & 12.4$\pm$1.8 & 0.36 \\   
A 1689 & a1689 & 0.018 & 10.4$\pm$0.115 & 0.41 & 1.52$\pm$0.07  & 12.1$\pm$1.8 & 0.27 \\   
A 401 & a401 & 0.099 & 8.6$\pm$0.071 & 0.36 & 1.45$\pm$0.09  & 9.5$\pm$1.7 & 0.25 \\   
A 2204 & a2204 & 0.057 & 8.5$\pm$0.035 & 0.42 & 1.39$\pm$0.08  & 9.0$\pm$1.6 & 0.30 \\   
A 2034 & a2034 & 0.015 & 8.3$\pm$0.107 & 0.31 & 1.40$\pm$0.09  & 8.8$\pm$1.6 & 0.22 \\   
A 1413 & a1413 & 0.018 & 8.3$\pm$0.107 & 0.36 & 1.38$\pm$0.08  & 8.7$\pm$1.6 & 0.26 \\   
A 2744 & a2744 & 0.014 & 8.7$\pm$0.229 & 1.09 & 1.30$\pm$0.08  & 8.6$\pm$1.5 & 0.84 \\   
A 1835 & a1835 & 0.020 & 8.5$\pm$0.040 & 0.42 & 1.32$\pm$0.08  & 8.5$\pm$1.5 & 0.32 \\   
PKS 0745-191 & pks0745 & 0.373 & 7.9$\pm$0.031 & 0.40 & 1.37$\pm$0.09  & 8.2$\pm$1.6 & 0.29 \\   
A 665 & a665 & 0.043 & 7.3$\pm$0.137 & 0.48 & 1.26$\pm$0.09  & 6.9$\pm$1.5 & 0.38 \\   
CYGNUS A & cygnusa & 0.272 & 6.9$\pm$0.022 & 0.29 & 1.30$\pm$0.10  & 6.8$\pm$1.5 & 0.22 \\   
ZW 3146 & zw3146 & 0.025 & 7.0$\pm$0.064 & 0.39 & 1.18$\pm$0.09  & 6.3$\pm$1.4 & 0.33 \\   
A 520 & a520 & 0.057 & 6.7$\pm$0.073 & 0.79 & 1.20$\pm$0.09  & 6.1$\pm$1.4 & 0.66 \\   
A 1795 & a1795 & 0.012 & 6.2$\pm$0.008 & 0.30 & 1.23$\pm$0.10  & 5.8$\pm$1.4 & 0.24 \\   
A 1650 & a1650 & 0.013 & 6.0$\pm$0.036 & 0.28 & 1.20$\pm$0.10  & 5.5$\pm$1.4 & 0.23 \\   
A 3667 & a3667 & 0.044 & 5.8$\pm$0.025 & 0.37 & 1.20$\pm$0.10  & 5.3$\pm$1.4 & 0.31 \\   
A 907 & a907 & 0.054 & 5.8$\pm$0.061 & 0.28 & 1.14$\pm$0.10  & 5.0$\pm$1.3 & 0.24 \\   
A 521 & a521 & 0.049 & 5.9$\pm$0.177 & 0.54 & 1.10$\pm$0.10  & 4.8$\pm$1.3 & 0.49 \\   
A 1995 & a1995 & 0.012 & 5.9$\pm$0.173 & 0.31 & 1.06$\pm$0.09  & 4.7$\pm$1.2 & 0.29 \\   
A 2146 & a2146 & 0.030 & 5.7$\pm$0.031 & 0.50 & 1.08$\pm$0.10  & 4.6$\pm$1.2 & 0.46 \\   
MS0735.6+7421 & ms0735 & 0.033 & 5.5$\pm$0.030 & 0.30 & 1.08$\pm$0.10  & 4.5$\pm$1.2 & 0.28 \\   
MS 1455.0+2232 & ms1455 & 0.032 & 5.1$\pm$0.055 & 0.31 & 1.01$\pm$0.10  & 3.9$\pm$1.2 & 0.30 \\   
A 2199 & a2199 & 0.009 & 4.4$\pm$0.010 & 0.19 & 1.05$\pm$0.12  & 3.5$\pm$1.2 & 0.18 \\   
A 496 & a496 & 0.038 & 4.3$\pm$0.012 & 0.20 & 1.04$\pm$0.12  & 3.4$\pm$1.2 & 0.19 \\   
A 2597 & a2597 & 0.025 & 4.0$\pm$0.014 & 0.21 & 0.98$\pm$0.12  & 2.9$\pm$1.1 & 0.21 \\   
3C348 (HERCULES A) & 3c348 & 0.062 & 3.9$\pm$0.032 & 0.23 & 0.94$\pm$0.12  & 2.7$\pm$1.1 & 0.25 \\   
A 1775 & a1775 & 0.010 & 3.7$\pm$0.047 & 0.22 & 0.94$\pm$0.13  & 2.6$\pm$1.1 & 0.23 \\   
HYDRA A & hydraa & 0.047 & 3.5$\pm$0.009 & 0.22 & 0.93$\pm$0.13  & 2.4$\pm$1.1 & 0.23 \\   
2A 0335+096 & 2a0335 & 0.175 & 2.9$\pm$0.005 & 0.18 & 0.84$\pm$0.15  & 1.8$\pm$1.0 & 0.21 \\   
SERSIC 159-03 & sersic159 & 0.011 & 2.8$\pm$0.012 & 0.17 & 0.82$\pm$0.15  & 1.7$\pm$0.9 & 0.20 \\   
A 2052 & a2052 & 0.027 & 2.3$\pm$0.002 & 0.15 & 0.76$\pm$0.17  & 1.3$\pm$0.9 & 0.19 \\   

\noalign{\smallskip}\hline
\end{tabular}
\tablefoot{
\tablefoottext{a}{Most commonly used cluster name and abbreviated catalogue names of clusters (compare to tables \ref{tab:cat1},\ref{tab:cat2}, and \ref{tab:cat3}). Sorted on descending mass ($\rm{M_{500}}$).}
\tablefoottext{b}{Maximum radius (from X-ray peak) covered in our analysis.}
\tablefoottext{c}{Overdensity radii $\rm{r_{500}}$ and $\rm{M_{500}}$ were calculated using the mass-temperature scaling relation from \citet{2009ApJ...692.1033V} for an estimate on the $\rm{r_{500}}$ fraction covered in each object. $\rm{<T~map>}$ was used as input for the scaling relation and errors were estimated assuming a 0.5\,keV systematic uncertainty (see scatter in Fig. \ref{fig:v09comp}).}
}
\end{center}
\label{tab:sampro}
\end{table*}
%--------------------------
We used archival observations with the \emph{Chandra} Advanced CCD Imaging Spectrometer \citep[ACIS,][]{2003SPIE.4851...28G} using the imaging (-I) or spectral (-S) CCD array (about 0.1 to 10\,keV energy range). This instrument provides high spatial ($\rm{\sim1\arcsec}$) and spectral resolution ($\rm{\sim100~eV}$ full width half maximum, FWHM). The field of view (FOV) is limited, so that only the inner $\rm{5-10\arcmin}$ of any cluster are homogeneously covered. See Sect. \ref{sec:aobs} for a list of all analysed clusters and their individual exposure times.

\subsection{Sample selection}
\label{sec:sample}

We based our sample selection on the NORAS \citep[378 sources, see][]{2000ApJS..129..435B}, REFLEX \citep[447 sources, see][]{2004A&A...425..367B}, and CIZA \citep[73 sources, see][]{2002ApJ...580..774E} catalogues. They all have been derived from \emph{ROSAT} observations, which deliver the only true imaging all-sky X-ray survey to date. NORAS and REFLEX cover the regions north and south of the galactic plane ($\rm{\pm 20^{\circ}}$), excluding the Magellanic Clouds. The CIZA sample covers the Galactic plane and thus adds some interesting clusters to our sample. However in the Galactic plane we have to deal with higher foreground absorption of X-rays (due to the column density $\rm{n_{H}}$ of the Galaxy).

Not all of these clusters have been observed with \emph{Chandra}, but predominantly X-ray bright clusters have, where the structure of the ICM could be well studied. We matched all \emph{Chandra} ACIS observations available from the \emph{Chandra} Data Archive (CDA\footnote{http://cxc.harvard.edu/cda/} on 2013-10-09) with cluster positions in the \emph{ROSAT} catalogues mentioned above.

\begin{figure}
  \resizebox{\hsize}{!}{\includegraphics[angle=0,trim=1cm 0cm 5cm 22cm,clip=true]{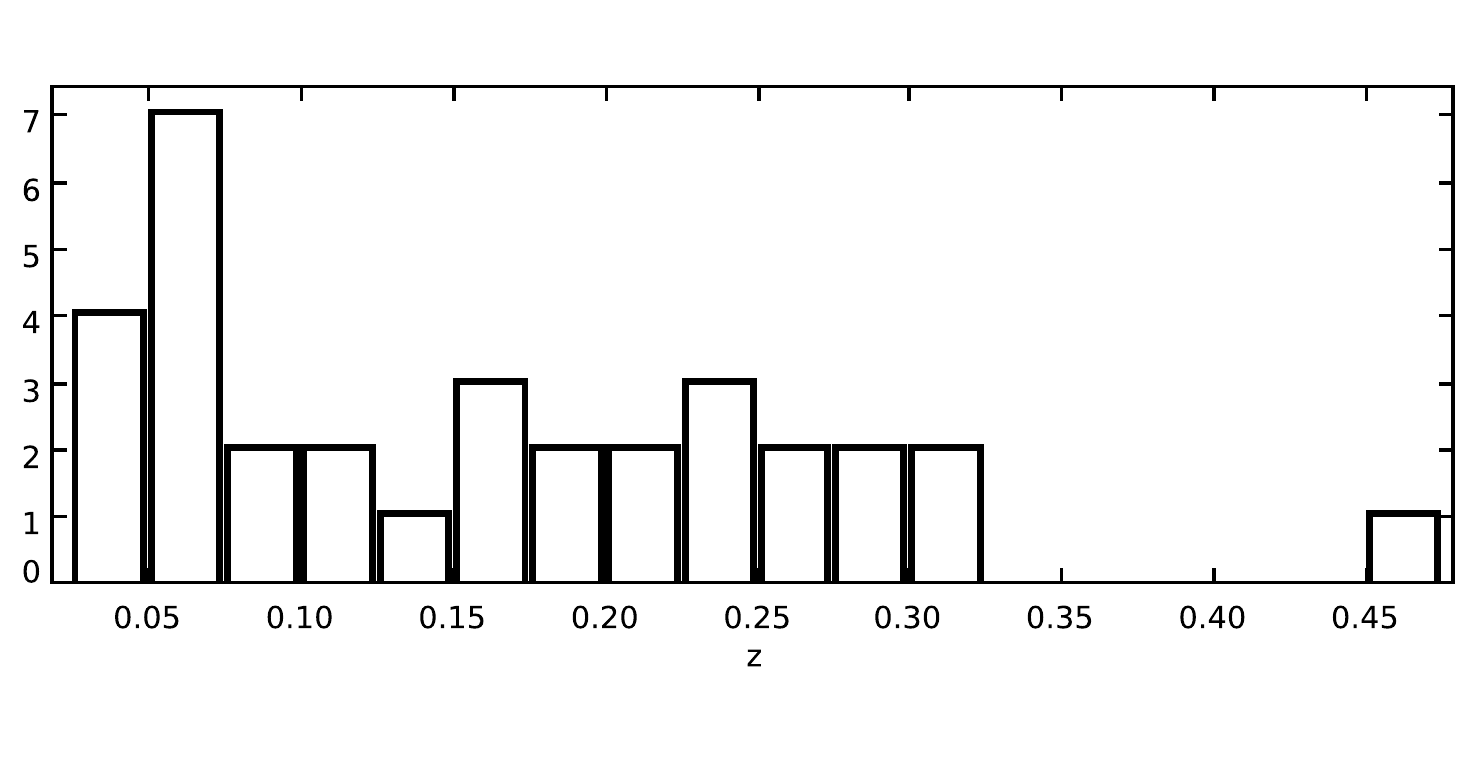}}
  \caption{
    Histogram of the redshift (z) distribution of the final sample of 33 clusters.}
  \label{fig:zhist}
\end{figure}

We set a luminosity cut on the $\rm{\sim300}$ clusters correlated with \emph{Chandra} observations, and only added clusters to the sample which had a luminosity of more than $\rm{2.0x10^{44} ~erg/s}$ in the 0.1-2.4\,keV \emph{ROSAT} energy band. We only accepted clusters and groups of galaxies with a redshift of $\rm{z \gtrsim 0.025}$ to ensure all clusters fit reasonably well into the \emph{Chandra} ACIS FOV. This excludes some nearby extended systems where larger radii are not homogeneously covered \citep[e.g. the Coma cluster, see][]{2009ApJ...692.1060V}. After these selections we used all clusters with $\rm{\gtrsim 100~ks}$ raw \emph{Chandra} exposure time. Our final sample consists of 33 X-ray bright, massive, nearby clusters of galaxies. The velocity dispersion of galaxies in the clusters is around $\rm{1000~km/s}$ for most systems. The cluster halo masses within the overdensity radius $\rm{r_{500}}$ range from $\rm{1\times10^{14}~M_{\odot}}$ to $\rm{2\times10^{15}~M_{\odot}}$ (Tab. \ref{tab:sampro}). At $\rm{r_{500}}$ the average density of the cluster is 500 times the critical density of the universe at the cluster redshift. The luminosity range is $\rm{(2-63)x10^{44}~erg/s}$ (0.1-2.4\,keV X-ray luminosity, see Tabs. \ref{tab:cat1}, \ref{tab:cat2}, and \ref{tab:cat3}), the redshift ranges from 0.025 to 0.45 (see Fig. \ref{fig:zhist}) and the total exposure analysed in this work is $\rm{\sim8~Ms}$ (corresponding to more than 90 days of observations). 
For a list of observations used in this study see Tab. \ref{tab:obsids}.

\subsection{Cluster maps}
\label{sec:maps}

\begin{figure*}
 \centering
 \resizebox{\hsize}{!}{\includegraphics[width=\linewidth,trim=0.5cm 5cm 0cm 1cm,clip=true]{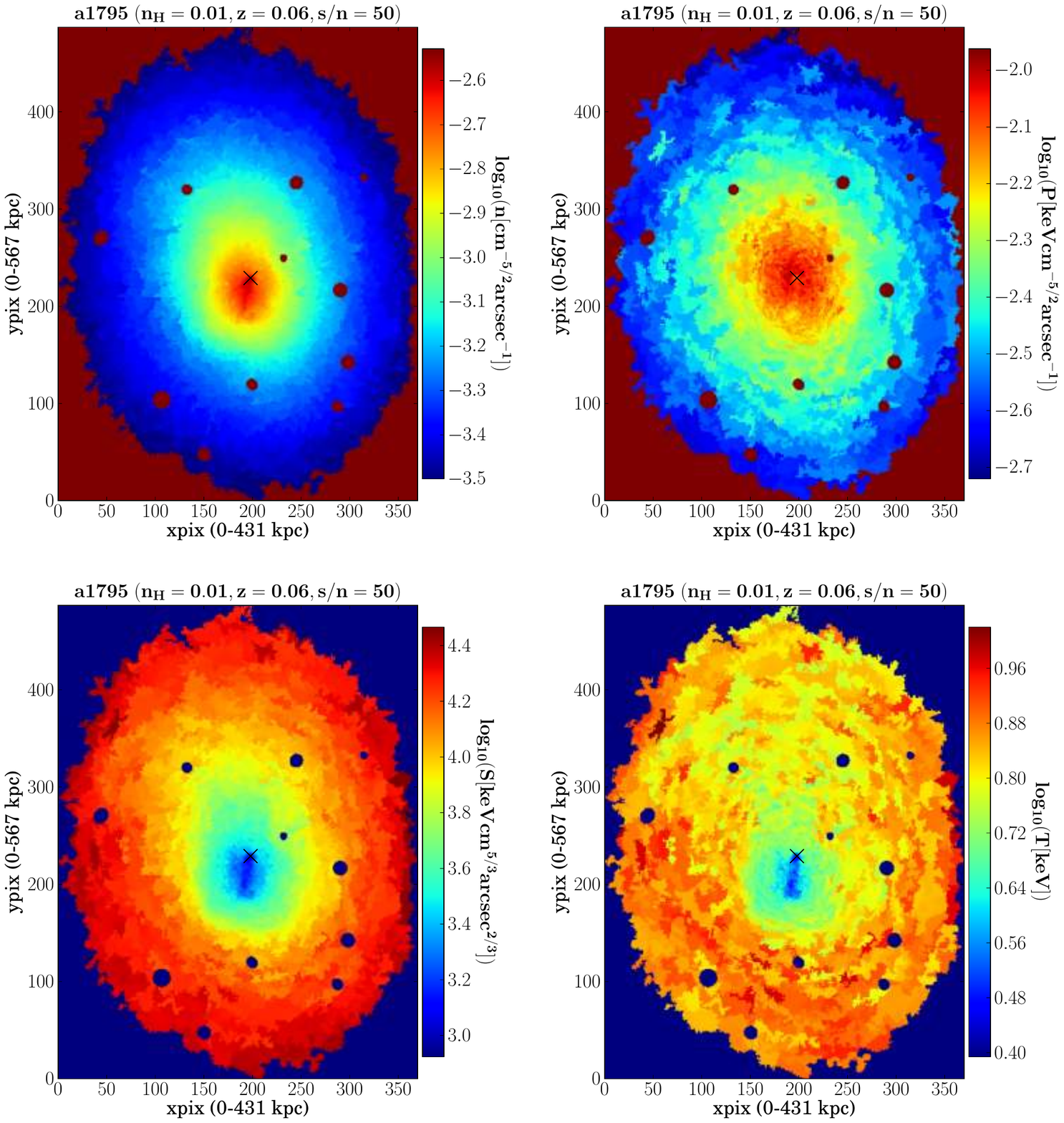}}
 \caption{
 From top left to bottom right: 2D maps of projected temperature, pressure, density and entropy of A\,1795. The cross marks the X-ray peak and centre for profile analysis. Point sources and regions below the surface brightness cut are set to zero. xpix and ypix are pixels along RA and DEC direction and the overall range in kpc is given. Scale: $\rm{1pix\sim1\arcsec}$. The plot titles indicate the abbreviated cluster name, the average foreground column density $\rm{n_H~[cm^{-2}]}$, and the redshift.}
 \label{fig:2dmaps}
\end{figure*}

For the \emph{Chandra} data reduction we used a pipeline of various \emph{Python} and \emph{C++} scripts to accurately map the structure of the ICM. The pipeline downloads all relevant datasets from the \emph{Chandra} data archive (CDA) and reprocesses them using the \emph{Chandra} standard data processing (SDP) with the \emph{Chandra} Interactive Analysis of Observations software package \citep[CIAO,][]{2006SPIE.6270E..1VF} version 4.5 and the \emph{Chandra} Calibration Database \citep[CalDB,][]{2007ChNew..14...33G} version 4.5.9.

A background light-curve for each observation is created and times of high background are removed from the event files, by iteratively removing times where the count rate is more than $\rm{3\sigma}$ from the median of the light-curve. Using the CIAO tool \texttt{acis\_bkgrnd\_lookup}, we find a suitable blank-sky background file, which is provided by the \emph{Chandra} X-ray centre \citep[see e.g.][]{2003ApJ...583...70M} and derive the background in each of the cluster observations. We correct residual spatial offsets between the individual observations if necessary by detecting point sources in each image with \texttt{wavdetect} and correlating the individual detection lists. Using the deepest observation as a reference, offsets in other event files are corrected by updating their aspect solution. With this procedure we ensure the best resolution of small-scale ICM variations. We create images from the event file of each dataset using an energy range of 0.5\,keV to 7.0\,keV and binning the image by a factor of two ($\rm{1pix\sim0.98\arcsec}$). For each image an exposure-map is created for an energy of 1.5\,keV. After the analysis of the individual observations, all images, background-, and exposure-maps are merged. A second point source detection with \texttt{wavdetect} is run on the merged images and after carefully screening the detection list, the point sources are masked from the image. In the following steps the image is adaptively smoothed with $snsmooth = 15$ and then binned into regions of equal S/N ratio of 50 ($\rm{>}$2500 counts per bin) using the contour binning technique \texttt{contbin} \citep[see][]{2006MNRAS.371..829S}. 

For the asymmetry (i.e. spread) analysis (Sect. \ref{sec:asym}) we generated maps with $\rm{S/N = 25}$ ($\rm{>}$625 counts per bin) for clusters where we obtained less than 50 independent spatial-spectral bins from the $\rm{S/N = 50}$ analysis. For each of the bins a detector response is calculated and the count spectrum extracted. Spectra of the same detectors are added together (ACIS-I and ACIS-S separately) and fitted using C-Statistics in \texttt{XSPEC version 6.11} \citep{1979ApJ...228..939C,1996ASPC...99..409A} with the \texttt{apec} model for collisionally-ionised diffuse gas, which is based on the ATOMDB code v2.0.2 \citep[][]{2012ApJ...756..128F}. The fit was done using a fixed foreground column density ($\rm{n_H~[cm^{-2}]}$, see Tab. \ref{tab:sampro}), which is determined from the Leiden/Argentine/Bonn (LAB) survey of Galactic HI \citep[based on][]{2005A&A...440..775K}, and a fixed redshift from the \emph{ROSAT} catalogues.

Only in the special case of PKS\,0745-191 was the fit done with $\rm{n_H}$ as a free parameter due to its location behind the Galactic plane and strong $\rm{n_H}$ variation within the FOV.

The free parameters of the fit to the count spectrum are temperature $\rm{T~[keV]}$, the metal abundance Z as a fraction of solar abundances \citep[reference solar abundance $\rm{Z_\odot}$ from][]{1989GeCoA..53..197A} and the normalisation $\rm{\eta~[cm^{-5}~arcsec^{-2}]}$ of the fit, which is defined as,

\begin{equation}
 \rm{\eta = 10^{-14}/(4\pi~{D_A}^2~(1+z)^2) ~\int{n_e~n_p~dV}}
\end{equation}

with $\rm{D_A}$ the angular diameter distance to the source, $\rm{n_e}$ and $\rm{n_p}$ the electron and hydrogen densities being integrated over the volume V. Assuming a spherical source, uniform density, and full ionisation with 10 per cent helium, 90 per cent hydrogen abundance (i.e. $\rm{n_e \sim 1.2 ~n_p}$) the hydrogen density can be calculated as

\begin{equation}
 \rm{n_p = \sqrt{\frac{(1+z)^2~10^{14}~\eta}{1.2\Theta^3~D_A}} \equiv \xi \cdot \sqrt{\eta}}
\end{equation}

where $\rm{\Theta}$ is the angular size of the source (see the ATOMDB webpage\footnote{http://atomdb.org/faq.php} for these equations). The factor $\rm{\xi}$ changes for different clusters and assumed geometries. We assumed $\rm{n_p \sim \sqrt{\eta}}$ within any given cluster.

The fit parameters of each bin are translated into images to obtain maps of the temperature, metal abundance and normalisation of the ICM.

From the 2D maps of the spectral fitting, we calculated a projected pseudo density, pressure, and entropy in each spatial bin of the ICM emission. We assumed a constant line-of-sight depth for all spectral regions calculating pseudo density n as square root of the fit normalisation, normalised by region size,

\begin{equation}
 \rm{n \equiv \sqrt{\eta}~[cm^{-5/2}~arcsec^{-1}]}
\end{equation}

pseudo pressure as,

\begin{equation}
 \rm{P \equiv n \times temperature~[keV~cm^{-5/2}~arcsec^{-1}]}
\end{equation}

and pseudo entropy as,

\begin{equation}
 \rm{S \equiv n^{-2/3} \times temperature~[keV~cm^{5/3}~arcsec^{2/3}]}
\end{equation}

We adopted a common definition of entropy for galaxy cluster studies, which is related to the standard definition of thermodynamic entropy s through 

\begin{equation}
 \rm{s = k_B~ln(S^{3/2}~(\mu~m_p)^{5/2})+s_0}
\end{equation}

with mean particle mass $\rm{\mu}$ and proton mass $\rm{m_p}$ \citep[see e.g.][]{2005RvMP...77..207V}. Relative cooling times of the ICM are proportional to $\rm{n^{-1} \times temperature^{1/2}~[keV^{1/2}~cm^{5/2}~arcsec]}$ where Bremsstrahlung emission dominates \citep[see e.g.][and references therein]{1986RvMP...58....1S}.

All distances were calculated using the redshift given in the \emph{ROSAT} cluster catalogues. All uncertainties are $\rm{1\sigma}$ confidence intervals unless stated otherwise. All further analysis only includes regions where the area-normalised normalisation of the fit was above $\rm{10^{-7}~cm^{-5}~arcsec^{-2}}$. This corresponds roughly to a surface brightness cut below which there were insufficient counts for detailed spectral analysis. For an example refer to the maps of Abell\,1795 in Fig. \ref{fig:2dmaps}. 

\section{Analysis of perturbations}
\label{sec:analyses}

Based on the very detailed spatial-spectral analysis of this sample of clusters with deep \emph{Chandra} observations we were able to study the ICM in great detail for a large sample of clusters. 

\subsection{Asymmetry measurement}
\label{sec:asym}

\begin{figure}
  \centering
  \includegraphics[width=\linewidth,trim=0cm 0.4cm 0cm 0.4cm,clip=true]{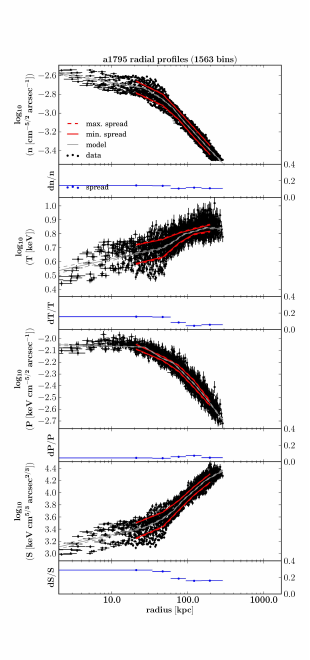}
  \caption{Radial profiles of projected density, temperature, pressure, and entropy of A\,1795. The centre is marked by a cross in Fig. \ref{fig:2dmaps}. Error bars are the fit-errors and the standard deviation of the radial distribution of the respective spatial-spectral bin. The Plotted lines show limits on intrinsic scatter around an average seven-node model (grey lines) within the given radial range (see Sect. \ref{sec:asym}). The small panels show the measured fractional scatter (M=5) with confidence- and radial-range.}
  \label{fig:a1795prof}
\end{figure}

One of the main goals of this study was to better characterise the thermodynamic state of the ICM in individual clusters and identify general trends in the whole sample. Important indicators of the state of the hot gas are fluctuations in thermodynamic properties. \citet{2013A&A...559A..78G} and \citet{2014A&A...569A..67G}, in recent high-resolution simulations, have shown a connection between such fluctuations and the Mach number of gas motion in the ICM. 

We examined the asymmetry (i.e. the spread) of thermodynamic properties in concentric, circular annuli of radius r around the peak X-ray emission. 
As input we used the S/N 50 ICM-maps (see Sect. \ref{sec:maps}). For nine clusters (a907, ms1455, a521, a665, a2744, a1775, a1995, 3c348, zw3146) we used S/N 25 instead to obtain at least five radial spread-bins with a minimum of five data points per bin for all clusters.
Assuming the data points are scattered statistically due to their uncertainties $\rm{\sigma_{stat}}$ we tested for intrinsic spread $\rm{\sigma_{spread}}$ in the radial profiles (see e.g. Fig. \ref{fig:a1795prof}). 

To avoid contamination of the spread-measurement by the slope, we modelled the radial profile by interpolating between several nodes. This modelled average profile can be described by a function $\rm{\mu(r,\mu_1,...,\mu_N)}$ with N=7. The nodes divide the cluster profile into bins with equal number of data points. We used a minimum of five data points per model-node and if this criterion was not met we reduced the number of nodes. 
The intrinsic fractional spread is given by a function of the form $\rm{\sigma_{spread}(r,\sigma_{spread,1},...,\sigma_{spread,M})}$.
We performed three independent spread measurements, splitting the profiles into one, two, and five radial bins (M=1,2,5) to measure the spread in the clusters with different radial resolutions. 

For every data point we obtained the mean value $\rm{\mu(r_i)}$ of the thermodynamic profile at its radius $\rm{r_i}$ by interpolating between the model-nodes (see grey lines in Fig. \ref{fig:a1795prof}). 
The individual spread values of each data point $\rm{\sigma_{spread}(r_i)}$ are constant within a given radial bin (see spread profile in Fig. \ref{fig:a1795prof}). 

The intrinsic spread of cluster properties was estimated using a Markov Chain Monte Carlo (MCMC) method implemented using \texttt{emcee} \citep[see][]{2013PASP..125..306F} with 100 walkers, 1000 iterations, and a burn-in length of 1000.
The total scatter of data values (i) was assumed to follow a Gaussian distribution with standard deviation of $\rm{\sigma_{tot}}$ calculated as,

\begin{equation}
 \rm{\sigma_{tot,i} = \sqrt{\sigma_{stat,i}^2 + \sigma_{spread}^2(r_i)}}
\end{equation}

For each iteration of the MCMC $\chi^2$ was calculated as,

\begin{equation}
 \rm{\chi^2 = \sum\limits_{i=1}^n \frac{(D_i-\mu(r_i))^2}{\sigma_{tot,i}^2}}
\end{equation}

where $\rm{D_i}$ are the individually measured values of the n data points (see Fig. \ref{fig:a1795prof}). The mean-model and spread-values have $\rm{N + M}$ free parameters. Using the MCMC method these parameters were varied, giving as probability value for each iteration, the logarithm of the likelihood (see e.g. \citet{2010arXiv1008.4686H}),

\begin{equation}
 \rm{log~L = -\frac{1}{2}~\chi^{2} -\frac{1}{2}~\sum\limits_{i=1}^n \log(2\pi~\sigma_{tot,i}^2})
\end{equation}

From this procedure we obtained a distribution of mean and spread values. We selected the best fit value for each parameter as the maximum of the distribution and estimated the uncertainty by giving the range containing 34 per cent of the obtained values on each side of the maximum. If the distribution was consistent with zero, we give the 68 per cent range as an upper limit. 

The spread measurements performed with just one radial spread-bin (M=1) were used to compare the overall fractional spread dn, dT, dP, and dS among clusters (see e.g. Figs. \ref{fig:dsdp}, \ref{fig:dtdn}, \ref{fig:dst}, and Tab. \ref{tab:persum}) and constrain the general ICM properties. 
For additional comparison of larger and smaller physical radii we measured the spread inside and outside of 100\,kpc from the X-ray peak (M=2, see e.g. $\rm{dP_{cen}}$ and $\rm{dP_{out}}$ Sect. \ref{sec:sam_inout}). From the spread analysis with five radial spread-bins (M=5) we obtained profiles of the intrinsic fractional standard deviation of thermodynamic properties dn/n, dT/T, dP/P, and dS/S (see Fig. \ref{fig:Proall}) in individual systems.
The spread measurements are consistent with the results of a second Monte Carlo simulation based technique, which was not based on Markov Chains \citep[similar to][]{2014MNRAS.444.1497S}.

\subsection{Surface brightness substructures}
\label{sec:sb}

The emissivity of the ICM and thus its surface brightness in X-rays is proportional to the plasma density squared \citep[for a review, see e.g.][]{1986RvMP...58....1S}. From the data reduction pipeline we obtained merged count and exposure images of the clusters. To remove any structure due to inhomogeneous exposure we divide the count image by the exposure image and obtain an image of the count rate. To identify small surface brightness fluctuations we enhanced the contrast of those images by unsharp-masking, a method commonly used in image analysis. This method is implemented by subtracting two versions of an image, smoothed by two different Gaussian filters, from each other. We subtracted an image smoothed with a Gaussian function of $\rm{\sigma = 2 pixel}$ width from an image smoothed with a Gaussian function of $\rm{\sigma = 5 pixel}$ width (see Fig. \ref{fig:umall}). The obtained unsharp-masked images enhance surface brightness features, complementing the analysis of substructure in thermodynamic properties, and highlighting disturbed systems.

\subsection{Average cluster temperatures}
\label{sec:temp}

\begin{figure}[]
  \centering
  \resizebox{0.85\hsize}{!}{\includegraphics[angle=0,trim=0 0cm 0cm 0cm,clip=true]{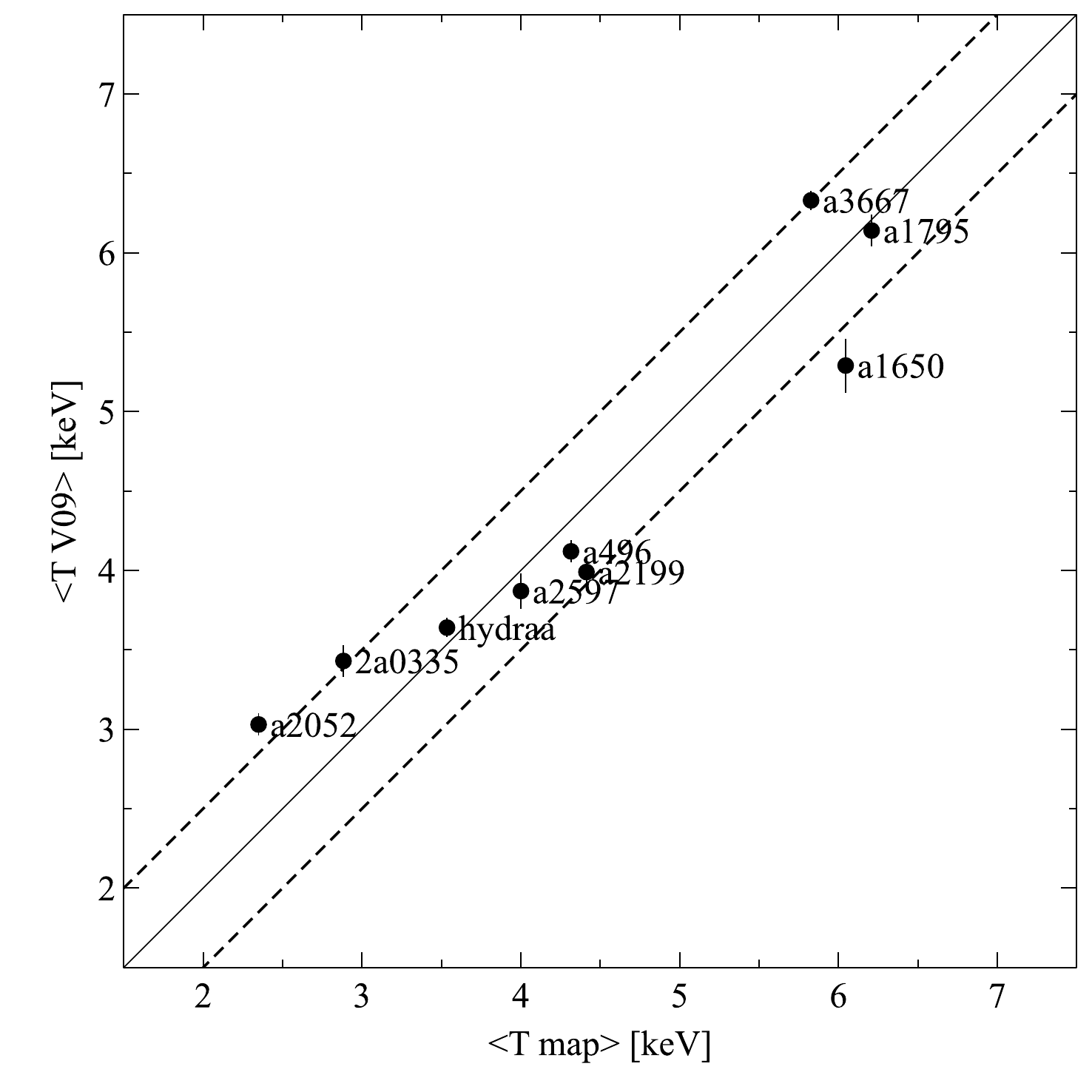}}
  \caption{
    Comparison of area- and error-weighted average 2D map temperatures $\rm{<T~map>}$ with overlapping low-z sample temperatures of V09 \citep[][]{2009ApJ...692.1033V} showing a scatter of about 0.5\,keV around the one-to-one relation.}
  \label{fig:v09comp}
\end{figure}
\begin{figure}[]
  \centering
  \resizebox{0.85\hsize}{!}{\includegraphics[angle=0,trim=0 0 0cm 0,clip=true]{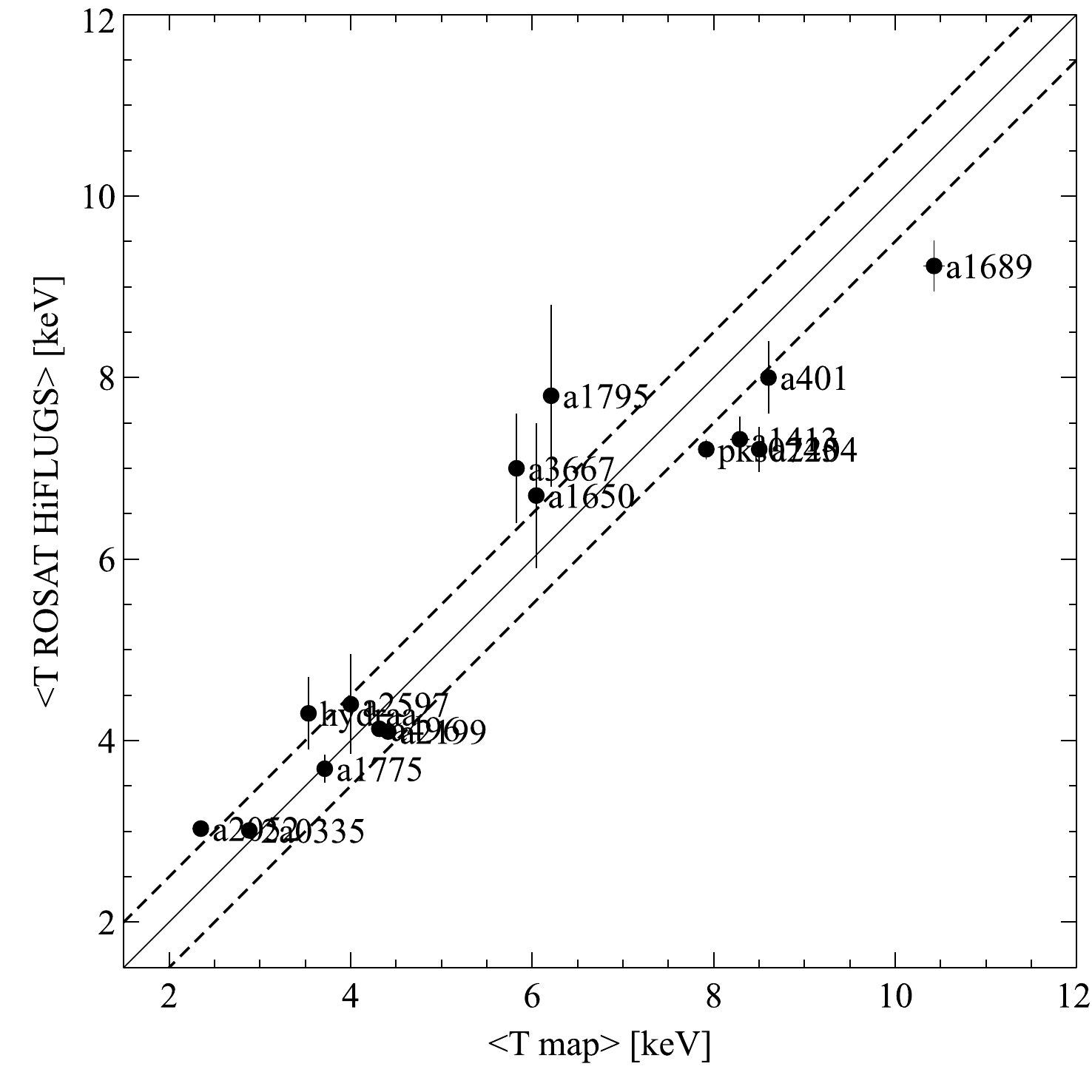}}
  \caption{
    Comparison of area- and error-weighted average 2D map temperatures $\rm{<T~map>}$ with HiFLUGS temperatures \citep[][]{2002ApJ...567..716R} from \emph{ROSAT} observations for the clusters overlapping with our sample. The dashed lines show a one-to-one relation with a scatter of 0.5\,keV.}
  \label{fig:Hicomp}
\end{figure}

We calculated average cluster temperatures $\rm{<T~map>}$ as the area- and error-weighted mean value of all measurements.
We computed the average of the bin temperature ($\rm{<T~prof>}$, see data points in Fig. \ref{fig:a1795prof}) which is usually lower than $\rm{<T~map>}$, because $\rm{<T~prof>}$ is weighted more on emission than area. Hotter regions in the outskirts are generally larger with lower emission and the relatively colder, X-ray bright, central regions cover a smaller area. 

We compared our approach for estimating overall cluster temperatures with previous studies like V09 (using \emph{Chandra}) or HiFLUGS (using \emph{ROSAT}) \citep[][]{2009ApJ...692.1033V,2002ApJ...567..716R} as shown in Fig. \ref{fig:v09comp} and Fig. \ref{fig:Hicomp}. Fig. \ref{fig:v09comp} shows significant scatter of up to 0.5\,keV between V09 and this study. This is expected since we use a different approach by averaging many independently fitted bins weighting by area and error-bars rather than by counts. The averaging of many different bins results in small error bars on our average temperature. Fig. \ref{fig:Hicomp} shows similar scatter when accounting for the larger error bars on the \emph{ROSAT} measurements. Note that we also use different extraction radii than the studies we compared to which influences the measured temperature.

To estimate the mass range of the cluster sample (see Tab. \ref{tab:sampro}) we used the mass-temperature scaling relation from \citet{2009ApJ...692.1033V} with the $\rm{<T~map>}$ temperatures as input to calculate the overdensity radius $\rm{r_{500}}$ and the total mass $\rm{M_{500}}$ included by this radius. We accounted for uncertainties in the scaling by assuming a systematic temperature uncertainty of 0.5\,keV (see scatter in Fig. \ref{fig:v09comp}). The average map temperature (area weighted) is comparable to the core-excised average temperature used for the scaling by \citet{2009ApJ...692.1033V}. 

\subsection{Table description}
\label{sec:tab}

\addtocounter{table}{+66} % needed if electronic tables included

The main results of the analysis of this study are summarized in separate tables for each cluster. We created 2D maps from the merged observations of every cluster and measured the asymmetries in thermodynamic parameters in circular concentric annuli around the centre.

\subsubsection{Map tables}
\label{sec:tab:1}

The primary data products of this analysis are 2D maps of the thermodynamic properties of the ICM based on the merged observations of ACIS-S and ACIS-I for every cluster in the sample (see Sect. \ref{sec:maps}).

{\bf Tables 2-34} contain the 2D map information (one table per cluster). Each table lists the properties for every pixel in the cluster maps ($\rm{1pix\sim0.98\arcsec}$). The map tables contain X (east-west direction) and Y (south-north direction) position (x, y, Cols. 1, 2) of the pixel, the index of the independently-fitted spatial-spectral bin it belongs to (binnum, Col. 3), the photon counts (cts, Col. 4), the background counts (bgcts, Col. 5), the effective exposure time (exp, Col. 6), the temperature and its upper and lower limits (T, Tup, Tlo, Cols. 7, 8, 9), the relative metallicity and limits (Z, Zup, Zlo, Cols. 10, 11, 12), the fit normalisation and limits (norm, normup, normlo, Cols. 13, 14, 15), the redshift (redshift, Col. 16), the foreground column density $\rm{n_H}$ (NH, Col 17), distance from the centre in pixels, arc seconds, and kpc (cen\_dist, rad\_arcsec, rad\_kpc, Cols. 18, 19, 20), the angle with respect to the west direction (cen\_angl, Col. 21), the calculated projected pressure and symmetric uncertainty (P, P\_err, Cols. 22, 23), projected entropy and uncertainty (S, S\_err, Cols. 24, 25), and density and uncertainty (n, n\_err, Cols. 26, 27). All uncertainties are on $\rm{1\sigma}$ confidence level.

\subsubsection{Asymmetry tables}
\label{sec:tab:2}

The secondary data products are based on the 2D maps and contain the measured spread (i.e. asymmetry, deviation from radial symmetry of the thermodynamic parameters, see M=5 in Sect. \ref{sec:asym}).

{\bf Tables 35-67} provide the measured properties in the concentric annuli for one cluster each. They contain the average radius of the annulus (rr, Col. 1), average bin-temperature $\rm{<T>}$ and uncertainty (T, Te, Cols. 2, 3), and average bin-metallicity and uncertainty (Z, Ze, Cols. 4, 5). For the intrinsic fractional spread values in projected properties best fit value and $\rm{1\sigma}$ upper and lower confidence limits are provided. Spread measurements contain the intrinsic spread in pressure with upper and lower limits (dP, dP\_eu, dP\_el, Cols. 6, 7, 8), entropy with limits (dS, dS\_eu, dS\_el, Cols. 9, 10 ,11), density with limits (dn, dn\_eu, dn\_el, Cols. 12, 13, 14), and temperature with limits (dT, dT\_eu, dT\_el, Cols. 15, 16, 17).

\section{Results}
\label{sec:results}

The detailed analysis of this sample of clusters enabled us to derive information on the thermodynamic state of the ICM in individual clusters and the whole sample in general.

\subsection{Perturbations in thermodynamic properties}
\label{sec:sam_turbobs}

\begin{figure}
  \resizebox{\hsize}{!}{\includegraphics[angle=0,clip=]{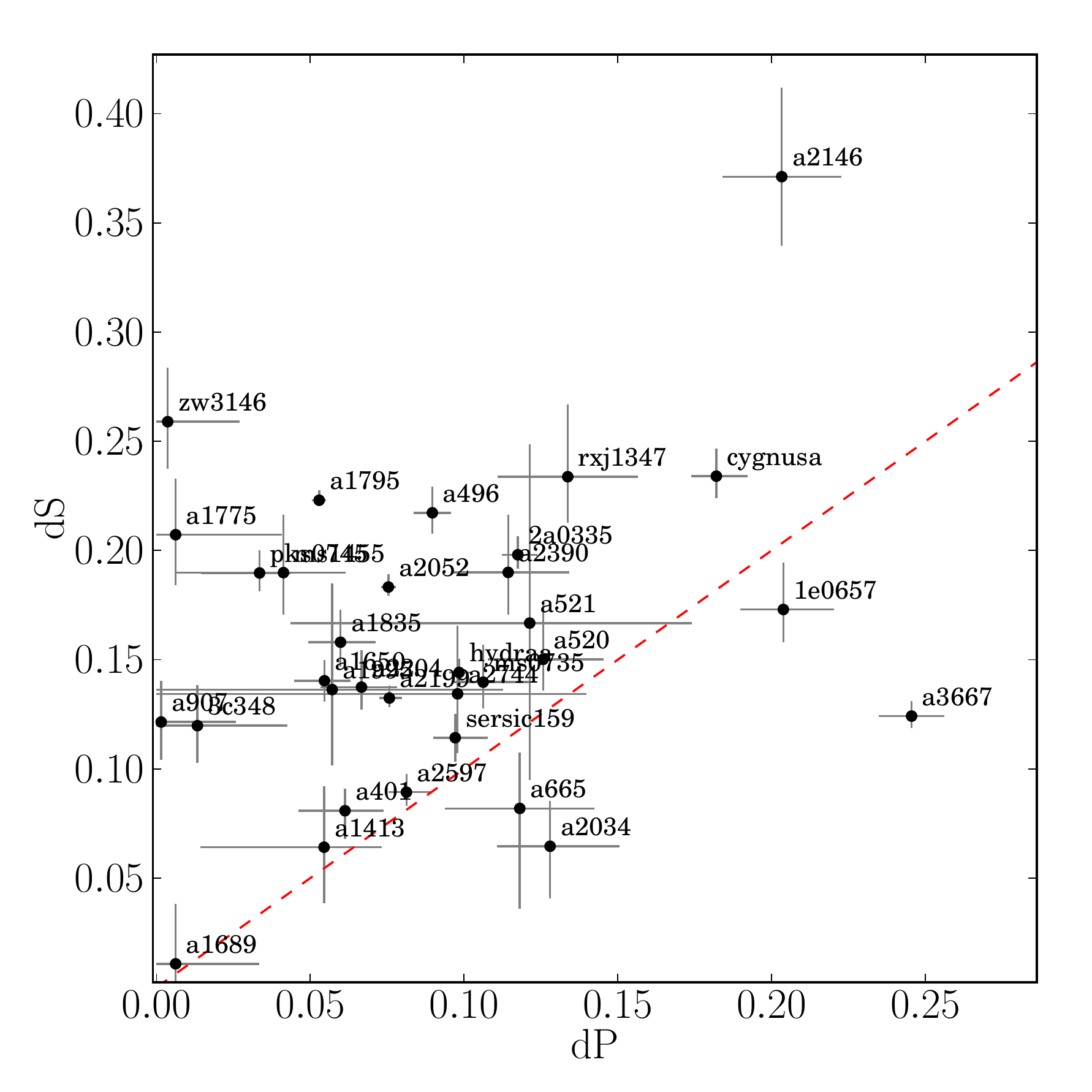}}
  \caption{
    Comparison of average projected pressure and projected entropy fluctuations for all clusters in the sample. The dashed line represents a one-to-one relation. The plot suggests that entropy fluctuations dominate in most clusters of the sample. Error bars are the statistical uncertainty from the MCMC measurements.}
  \label{fig:dsdp}
\end{figure}

\begin{figure}
  \resizebox{\hsize}{!}{\includegraphics[angle=0,clip=]{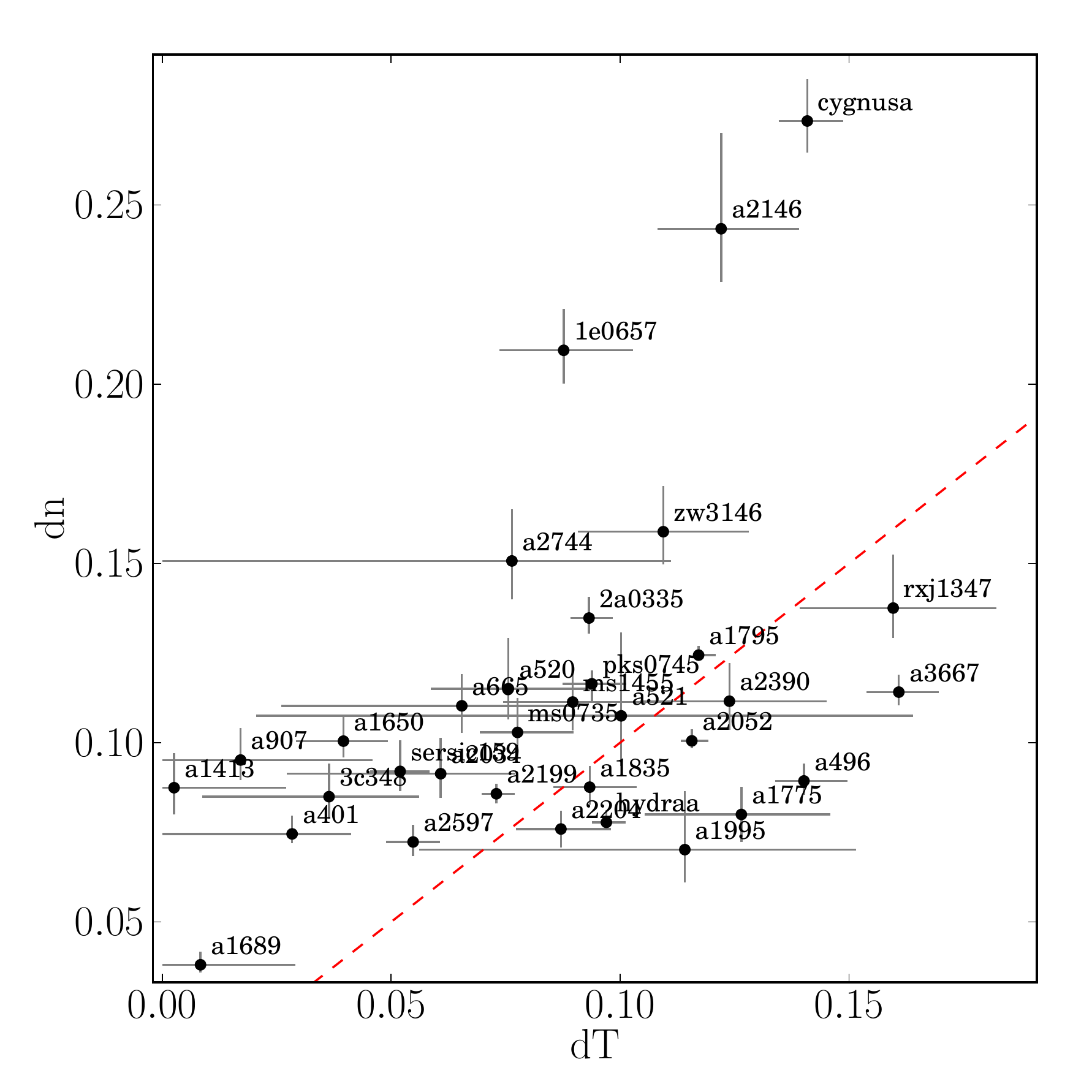}}
  \caption{Comparison of average temperature and projected density fluctuations for all clusters in the sample. The dashed line represents a one-to-one relation. Error bars are the statistical uncertainty from the MCMC measurements.}
  \label{fig:dtdn}
\end{figure}

We studied the average fluctuations in thermodynamic properties of all clusters in the sample using the M=1 spread calculations from Sect. \ref{sec:asym}. Comparing the average measurements for all systems in the sample enabled us to find general trends.

Fig. \ref{fig:dsdp} indicates that, on average, entropy fluctuations (fractional spread, dS) dominate pressure fluctuations (fractional spread, dP). For some clusters we only obtained upper limits (e.g. A\,1689). There are some outliers (e.g. A\,2146 and A\,3667), which are heavily disturbed systems (see Sect. \ref{sec:sam_turbcav}). Overall most clusters lie off a one-to-one correlation with an average of 16 per cent fluctuations in entropy and 9 per cent in pressure (see Tab. \ref{tab:samper}).

Fig. \ref{fig:dtdn} shows a smaller offset between the average density fluctuations (fractional spread, dn) and temperature fluctuations (fractional spread, dT). We measure significant spread in dn for every cluster in the sample but only obtain upper limits on the dT spread for some. There are some outliers which show very strong density fluctuations (e.g. A\,2146, 1E\,0657-56, and Cygnus\,A). 
Overall most clusters are close to a one-to-one correlation with an average of 11 per cent fluctuations in density and 9 per cent in temperature (see Tab. \ref{tab:samper}).

%--------------------------
\begin{table}[]
\caption[]{Average perturbations.}
\begin{center}
\begin{tabular}{rrrr}
\hline\hline\noalign{\smallskip}
  \multicolumn{1}{l}{$\rm{<dP>}$\tablefootmark{a}} &
  \multicolumn{1}{l}{$\rm{<dS>}$\tablefootmark{a}} &
  \multicolumn{1}{l}{$\rm{<dn>}$\tablefootmark{a}} &
  \multicolumn{1}{l}{$\rm{<dT>}$\tablefootmark{a}} \\
\noalign{\smallskip}\hline\noalign{\smallskip}
 0.09 $\pm$ 0.06 & 0.16 $\pm$ 0.07 & 0.11 $\pm$ 0.05 & 0.09 $\pm$ 0.04 \\  
\noalign{\smallskip}\hline
\end{tabular}
\tablefoot{
\tablefoottext{a}{Average fractional perturbations in thermodynamic properties of the 33 sample clusters. Standard deviation as confidence range.}
}
\end{center}
\label{tab:samper}
\end{table}
%--------------------------

\subsection{Perturbations on different scales}
\label{sec:sam_inout}

\begin{figure}
 \centering
 \resizebox{0.88\hsize}{!}{\includegraphics[angle=0,trim=0cm 0.45cm 0.5cm 0.55cm,clip=true]{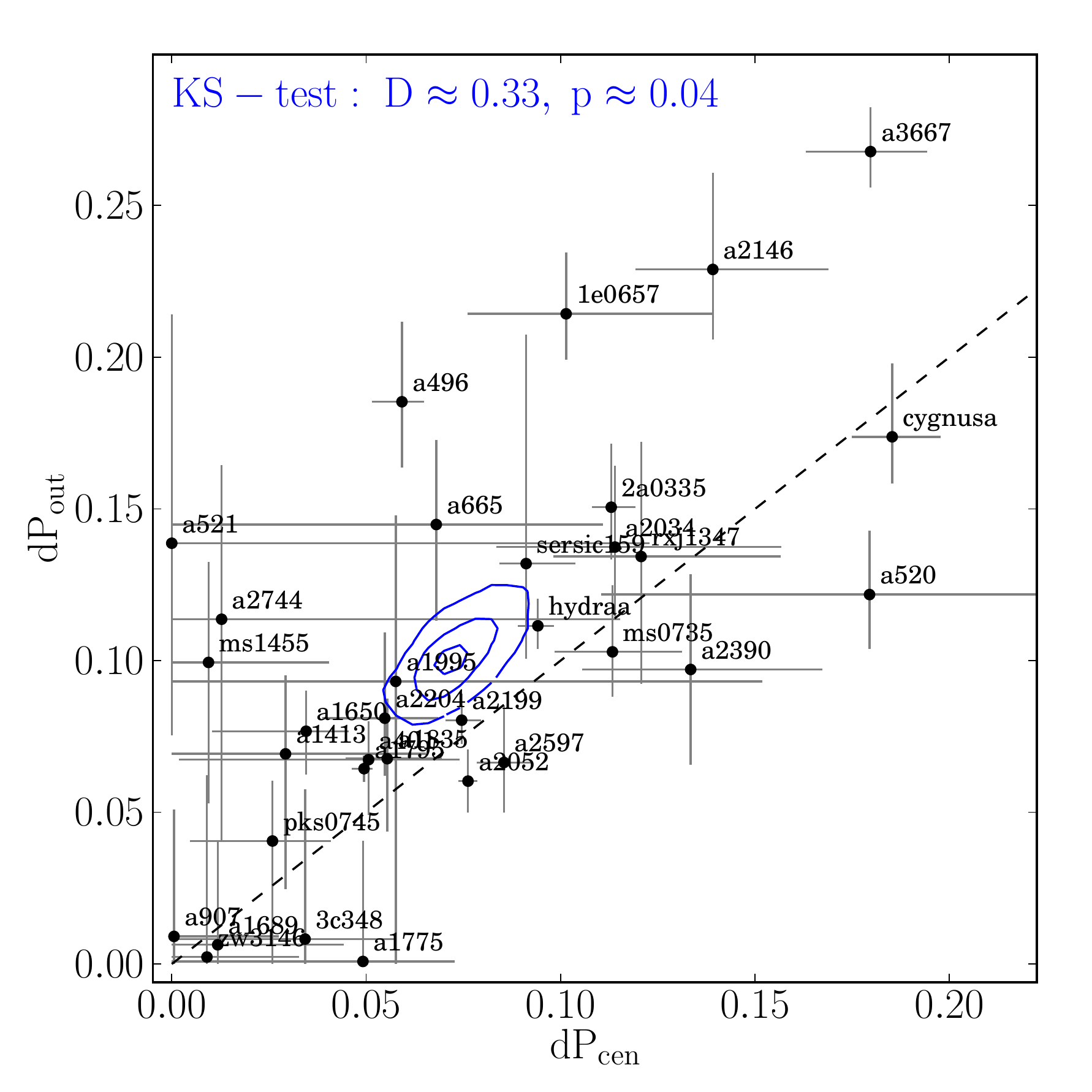}}
 \resizebox{0.88\hsize}{!}{\includegraphics[angle=0,trim=0cm 0.45cm 0.5cm 0.55cm,clip=true]{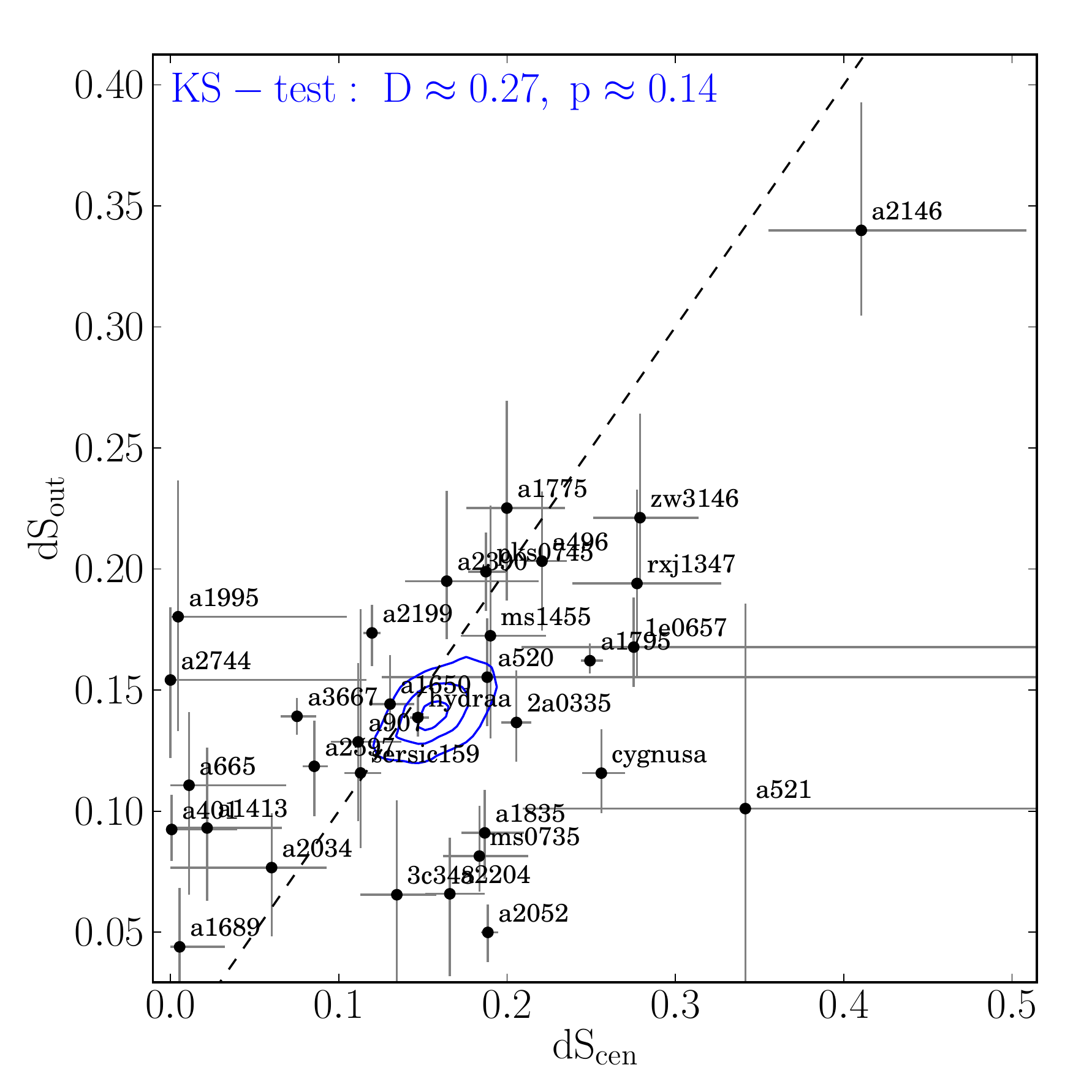}}
 \resizebox{0.88\hsize}{!}{\includegraphics[angle=0,trim=0cm 0.45cm 0.5cm 0.55cm,clip=true]{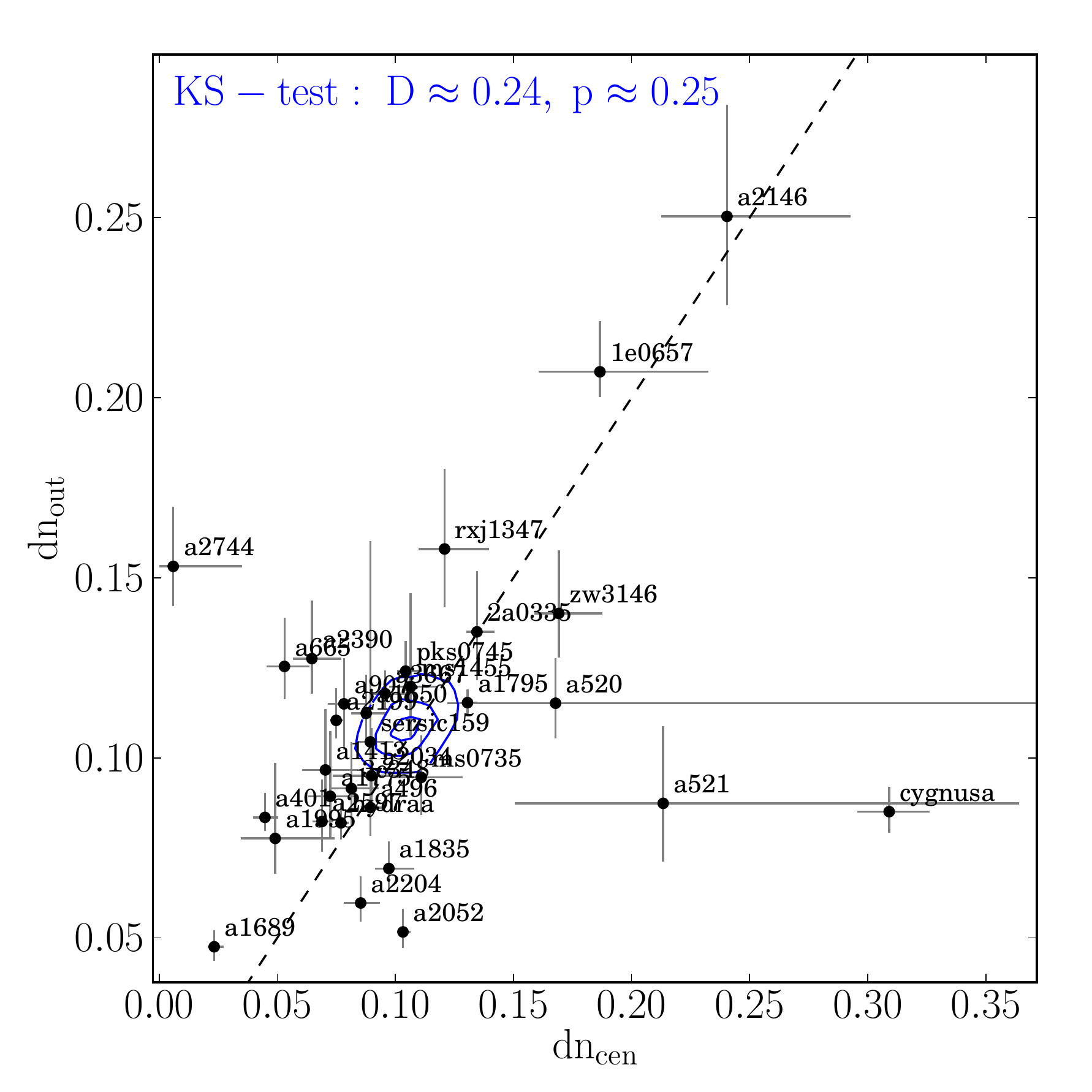}}
 \caption{Comparison of pressure- (top), entropy- (middle), and density-perturbations (bottom) in the central $\rm{\lesssim100~kpc}$ and outer $\rm{\gtrsim100~kpc}$ regions. Confidence regions of the mean value as contours (85, 50, and 15 per cent of peak value). KS-test results on top. One-to-one correlation as dashed line.}
 \label{fig:inout}
\end{figure}

We found evidence for larger pressure perturbations dP at larger scales. Fig. \ref{fig:inout} shows a clear separation of the distribution for central regions $\rm{\lesssim100~kpc}$ (cen) from the centre and regions beyond (out). 
To compare the distribution of spread values on different scales we used the asymmetry measurements from Sect. \ref{sec:asym} where the cluster profile is divided at $\rm{\sim100~kpc}$. 
The division at 100\,kpc was chosen, after visual inspection, as a robust separation radius between central substructure and outer more homogeneous regions. We did not choose the regions relative to $\rm{r_{500}}$ to ensure we are testing the same physical scale of the ICM fluctuations.

To estimate the mean of the distributions we used a bootstrapping re-sampling technique, calculating the mean value for 1000 permutations with repetition. The contours are at 15, 50, and 85 per cent of the maximum of the obtained distribution of mean values. In addition Fig. \ref{fig:inout} contains the output of a Kolmogorov-Smirnov (KS) test to quantify the difference between the cen and out distributions. The D value states the maximum fractional offset between the cumulative distribution graphs and the p value is the probability of the null hypothesis. This means the probability that the dP distributions are different is 96 per cent and thus just above the $\rm{2\sigma}$ level. It should be noted that the offset in dP is dominated by low dP data points and decreases to about $\rm{1\sigma}$ level when only including data points above 0.05\,dP. For the thermodynamic properties dS, dn, and dT there is no significant difference between inner and outer radii.

\subsection{Metallicity correlations}
\label{sec:sam_met}

\begin{figure}
 \centering
 \resizebox{0.95\hsize}{!}{\includegraphics[angle=0,clip=]{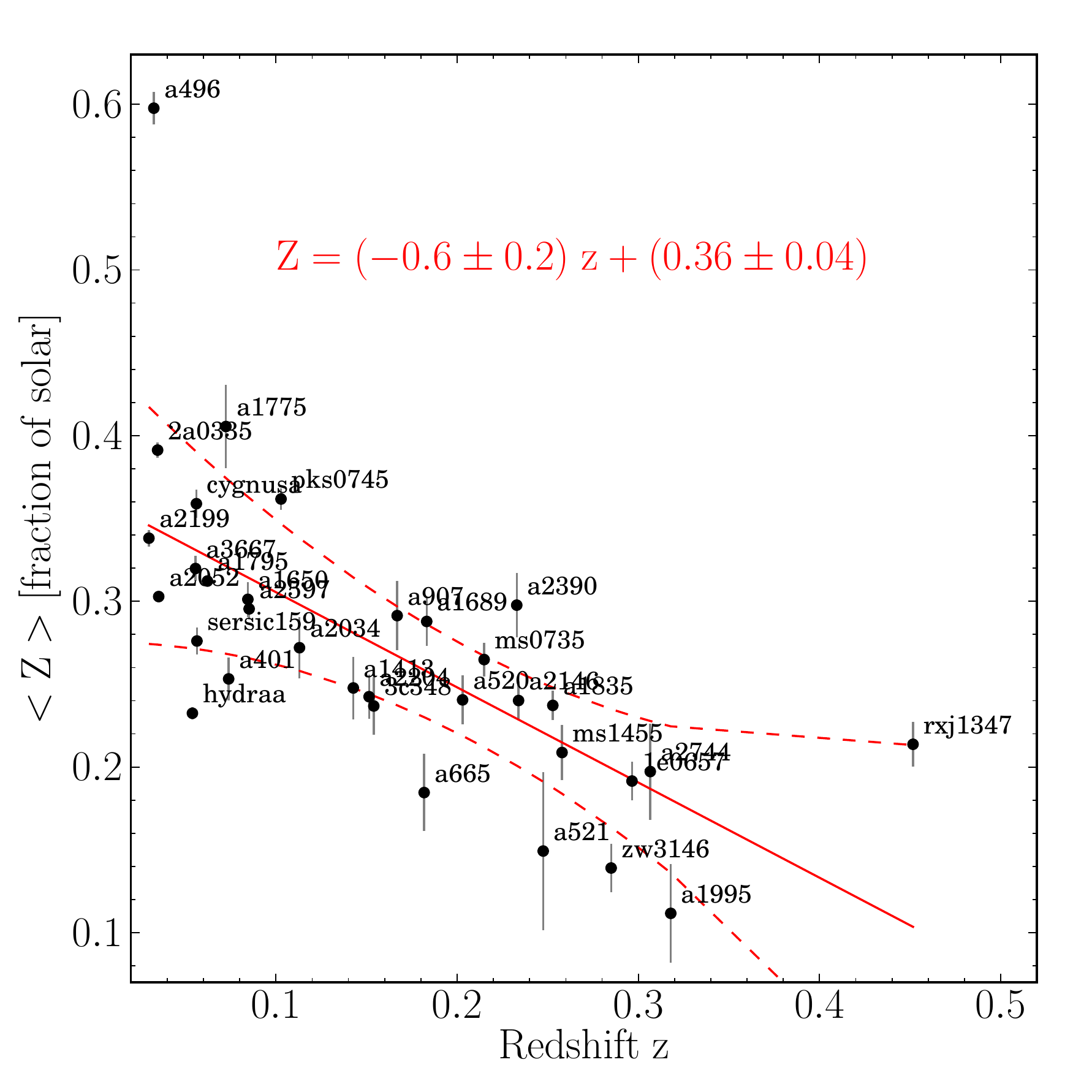}}
 \caption{Comparison of the cluster redshift z and the area- and error-weighted average 2D map metallicity measured in the ICM (full radial range). The red line and equation show the best fit linear correlation. Dashed lines indicate the $\rm{1\sigma}$ scatter around the best fit. Error bars are the statistical uncertainty of the weighted average.}
 \label{fig:metall}
\end{figure}

We found an anti-correlation between the average temperatures ($\rm{<T~map>}$, see Sect. \ref{sec:temp}) of the clusters and their average metallicity. The best fit linear correlation is $\rm{Z/Z_\odot = -(1.0\pm0.7)~T/100keV + (0.34\pm0.06)}$.
The average metallicity of the clusters have been weighted by area and error in the same way as the average map temperatures in Sect. \ref{sec:temp}. A similar correlation can be found within the individual clusters (see maps in Sect. \ref{sec:tab:1}). By repeating the same slope analysis for the inner ($\rm{\lesssim100~kpc}$) and outer ($\rm{\gtrsim100~kpc}$) regions of the cluster we found that the average metallicity is higher in the central regions and the slope of the T-Z anti-correlation is steeper. Testing different weighting methods for the average temperature and metallicity we obtained consistent correlations with some scatter.
Temperature is also correlated to the redshift of the clusters in this sample (more luminous and massive systems at higher redshift due to selection bias). 

We investigated the redshift-metallicity (z-Z) relation and found the correlation with redshift to be tighter than with temperature (see Fig. \ref{fig:metall}). The best fit linear correlation of metallicity and redshift is $\rm{Z/Z_\odot = -(0.6\pm0.2)~z + (0.36\pm0.04)}$ (see Fig. \ref{fig:metall}). There was no evidence that the z-Z anti-correlation is steeper in the central regions.
The average metallicity of the sample is $\rm{Z\approx0.3\pm0.1~Z_\odot}$. 
The significance of a correlation was estimated using a re-sampling technique, performing a linear fit on random sub-samples of 17 clusters and using the mean value and $\rm{1\sigma}$ width as best fit and error range of slope and normalisation.

\section{Individual clusters}
\label{sec:indi}

In addition to the sample properties the data contain important information on the properties of individual systems. The sample consists of clusters with a wide range of different structures from more relaxed systems like A\,496 and A\,2199 to disturbed clusters like the 1E\,0657-56 and A\,2146. We highlight some special cases below. Temperature maps, unsharp-masked images, and radial profiles of thermodynamic properties can be found in Sect. \ref{sec:amaps}, \ref{sec:amask}, and \ref{sec:arad}.\\
\\
{\bf Abell\,1795} has the deepest exposure in our sample and thus the most detailed spatial-spectral information on the ICM could be derived in this system. \citet{1994AJ....107..857O} found that its central cD galaxy has a peculiar radial velocity of $\rm{150~km~s^{-1}}$ within the cluster, while they measured the velocity dispersion of cluster members at $\rm{\sigma=920~km~s^{-1}}$. According to \citet{2001MNRAS.321L..33F} the simplest explanation for the visible soft X-ray filament would be a cooling-wake behind the cD galaxy (approximate position at the cross in Fig. \ref{fig:2dmaps}, filament extending to the south), which is oscillating in the DM potential of the galaxy cluster. The centre is surrounded by linear surface-brightness features which might be remnants of past mergers with subhalos or created by the outburst of a strong AGN \citep[see e.g.][]{2001ApJ...562L.153M, 2002MNRAS.331..635E, 2014MNRAS.445.3444W, 2015ApJ...799..174E}. 

The deep unsharp-masked count images in this study show the central surface-brightness features at higher significance than previous studies (see Fig. \ref{fig:umall}). 
The maps (Fig. \ref{fig:2dmaps}) contain 1563 bins with a S/N count ratio of 50. The detailed radial profiles of the projected thermodynamic properties (Fig. \ref{fig:a1795prof}) were the basis for the radial asymmetry measurements. Entropy perturbations seem to dominate throughout the cluster, suggesting that the 1D Mach number is comparable to the variance of entropy (see below, Sect. \ref{sec:sam_turbturb}). 
The perturbation measurements are influenced by the presence of a cooler X-ray filament (inner $\rm{\sim40~kpc}$) and projection-effects (see Sect. \ref{sec:sam_turbcav}).
The central filament seems to increase the spread in density, temperature, and entropy, but pressure seems unaffected (see scatter in Fig. \ref{fig:a1795prof}). The profiles follow the average trend of higher pressure perturbations in the outskirts (see Sect. \ref{sec:sam_inout}).
The findings confirm and improve the detection of many surface-brightness features at the centre of the cluster. The detailed analysis of thermodynamic perturbations support a model where isobaric processes dominate the central ICM.\\
\\
{\bf 1E\,0657-56} (the Bullet Cluster) offers an almost edge-on view of two massive merging subclusters \citep[][]{2002ApJ...567L..27M}. It was possible to find a significant offset between the total mass profile from weak-lensing and the X-ray emission of the  hot ICM and thus make a very convincing case for the existence of DM \citep[][]{2004ApJ...606..819M, 2004ApJ...604..596C, 2006ApJ...648L.109C, 2008ApJ...679.1173R}. 

We found the strongest surface-brightness fluctuations around the prominent Mach cone from the impact of the smaller subcluster (see Fig. \ref{fig:umall}). 
The density fluctuations in the cluster are among the highest in the observed sample (see Fig. \ref{fig:dtdn}). Pressure fluctuations around the ``Bullet'' are significantly weaker than at larger radii (see Fig. \ref{fig:Proall}). It should be noted that major merger shocks and AGN feedback are not modelled in the \citet{2014A&A...569A..67G} simulations and thus in those cases the direct connection between Mach number and fluctuations in thermodynamic parameters might change (see Sect. \ref{sec:sam_turbcav}). 1E\,0657-56 has the second highest $\rm{<T~map>}$ in the sample (after RX\,J1347-114) and the lowest average metallicity (see Fig. \ref{fig:metall}).
The Bullet Cluster has a particularly flat pressure profile with large scatter when compared to other clusters. In the most disturbed systems the pressure profile is flat and does not drop with radius.\\
\\
{\bf Abell\,2052} hosts an extended region of colder ICM at its centre caused by rising colder gas due to strong AGN feedback from the central cD galaxy. The radio source connected to the central AGN and its effect on the surrounding ICM have been studied in detail by \citet{2001ApJ...558L..15B, 2003ApJ...585..227B, 2011ApJ...737...99B}. Feedback from the radio source pushes the X-ray emitting gas away from the centre and creates a sphere of enhanced pressure around the central region. \citet{2015MNRAS.447.2915M} investigated different merger scenarios in recent simulations of the cluster. 

In our measurements the cluster emission shows a large scale ellipticity which could be an indicator of a past merger.
The cluster has the lowest average temperature in the sample ($\rm{<T~map>\sim2.3\,keV}$, see Tab. \ref{tab:sampro}) and one of the largest drops in entropy asymmetry from the centre to the outskirts (see Fig. \ref{fig:inout}).
At about 20\,kpc from the central AGN we detect an enhancement in projected pressure of more than a factor of two compared to the enclosed ICM (see Fig. \ref{fig:Proall}). On larger scales the cluster shows a spiral structure in surface brightness which is most likely caused by sloshing of gas due to a past merger (see Fig. \ref{fig:umall}).
In A\,2052 all thermodynamic profiles are heavily influenced by the AGN feedback at the center. Only the strongest feedback cases in our sample show a deviation from a radially decreasing pressure profile.\\
\\
{\bf Abell\,2146} is a major merger viewed almost edge-on. \citet{2010MNRAS.406.1721R, 2011MNRAS.417L...1R, 2012MNRAS.423..236R} detected two opposing shock fronts in the cluster and investigated transport processes in the ICM. They find the system to be less massive and thus colder than the Bullet Cluster merger (see also Fig. \ref{fig:dst}). Unlike in the Bullet Cluster, the secondary BCG in A\,2146 seems to be slightly lagging behind the shockfront \citep[][]{2012MNRAS.420.2956C}. 

The cluster shows the highest entropy perturbations in the sample (see Fig. \ref{fig:dsdp}). 
Asymmetries are especially high around the cluster centre which we choose on the smaller merging subcluster (see the cross in Fig. \ref{fig:Tall}). Like in 1E\,0657-56, pressure fluctuations around the merging subcluster are rather low but increase at larger radii (see Fig. \ref{fig:inout}). The difference in dP/dS between outer and inner radii is among the highest in the sample (see Fig. \ref{fig:inout}).
The extreme entropy perturbations in A\,2146 indicate the highest average Mach number in our sample (see Sect. \ref{sec:sam_turbturb}), almost twice as high as in the Bullet Cluster. If the two merging systems have similar impact velocities, the lower temperature of A\,2146 could cause the Mach number of the turbulence induced by the merger to be twice as high as in the Bullet Cluster.\\
\\
{\bf Cygnus\,A and Hydra\,A} are two similar systems with strong AGN feedback. Both sources have strong radio jets emerging from the central AGN causing complex structures in the X-ray emitting ICM around the nucleus \citep[see e.g.][for Hydra\,A and Cygnus\,A respectively]{2000ApJ...534L.135M,2002ApJ...565..195S}. 
\citet{2002ApJ...568..163N,2005ApJ...628..629N} found AGN feedback to influence the ICM on large scales in Hydra\,A.

The unsharp-masked analysis of the surface-brightness images clearly shows the strong feedback structures around the central AGN (Fig. \ref{fig:umall}).
Average perturbations and temperature in Cygnus\,A are significantly higher than in Hydra\,A (see Figs. \ref{fig:dsdp}, \ref{fig:dst}). Cygnus\,A shows enhanced density and temperature at larger radii to the north-west (see Fig. \ref{fig:Tall}). Hydra\,A has a very asymmetric temperature distribution which is mainly caused by continuous radial structures of colder gas extending from the centre (see Fig. \ref{fig:Tall}).
The average metallicity of Cygnus\,A is significantly higher than for Hydra\,A (see Fig. \ref{fig:metall}). 
The thermodynamic profiles of Cygnus\,A are more disturbed which would indicate a stronger or more recent AGN outburst.\\
\\
{\bf Abell\,2199} is a typical relaxed cluster with a cool core and AGN feedback structures at its centre \citep[][]{1999ApJ...527..545M,2002MNRAS.336..299J}. \citet{2013ApJ...775..117N} found various substructures in deep \emph{Chandra} observations of the cluster, including evidence for a minor merger $\rm{\sim400~Myr}$ ago. \citet{2006MNRAS.371L..65S} found a weak isothermal shock ($\rm{\sim100kpc}$ from the centre to the south-east) likely caused by the supersonic inflation of radio lobes by jets from the central AGN (3C\,338). 

Our maps indicate a weak temperature jump in the area where the shock has been detected (see Fig. \ref{fig:Tall}) and show a large scale asymmetry in the temperature distribution between north and south (also visible in the scatter of the radial profile, see Fig. \ref{fig:Proall}). The surface brightness unsharp-masked image shows some of the structures at the centre and a weak indication of the surface brightness jump due to the shock to the south-east. 
The perturbations in entropy, pressure, temperature, and density are in the average regime of our sample as expected for an overall relaxed system (see Figs. \ref{fig:dsdp}, \ref{fig:dtdn}).
The radial profiles show a prominent discontinuity around 50\,kpc from the center, related AGN feedback. Just outside this jump there is a region where the fit uncertainties are larger due to overlapping chip gaps of many observations, which could bias our results (see faint spurious linear structures in Fig. \ref{fig:Tall}).\\
\\
{\bf Abell\,496} is a relaxed cluster with relatively high metallicity around the cool core and non-uniform temperature distribution on large scales \citep[][]{2001A&A...379..107T,2006PASJ...58..703T,2014A&A...570A.117G}. By comparing dedicated simulations of the cluster to deep \emph{Chandra} observations \citet{2012MNRAS.420.3632R} concluded that the cluster was most likely perturbed by a merging subcluster 0.6-0.8\,Gyr ago.

The spiral surface brightness excess structure and the northern cold front is clearly visible in our new images of temperature and unsharp-masked count images (see Fig. \ref{fig:Tall}, \ref{fig:umall}). 
The radial profiles and asymmetry measurements show relatively large spread in entropy (see Fig. \ref{fig:Proall}). Figs. \ref{fig:dsdp} and \ref{fig:dtdn} show that the cluster has a large average temperature spread (due to the large scale asymmetry caused by the sloshing of colder gas from the centre) and thus also larger spread in entropy.
A\,496 has a flat pressure profile similar to the strongest merging systems in the sample. The average metallicity is the highest we measured.\\
\\
{\bf PKS\,0745-191} is a relaxed cluster at large scales out to the virial radius \citep[][]{2009MNRAS.395..657G}. \citet{2014MNRAS.444.1497S} found AGN feedback and sloshing structures in deep \emph{Chandra} observations. 

The AGN feedback features are most prominent in the unsharp-masked image (Fig. \ref{fig:umall}) and indication of weak asymmetry in temperature on large scales can be seen in Fig. \ref{fig:Tall}. The radial profiles of the cluster follow the expected trend for relaxed systems but show remarkably low intrinsic scatter in pressure and relatively high scatter in entropy (see Fig. \ref{fig:Proall}). The scatter measurements of \citet{2014MNRAS.444.1497S} are consistent with our method for dn, dT, dP, and dS. This is the only cluster in the sample where $\rm{n_H}$ was set as a free parameter in the spectral fits, because the foreground column density of this system is knows to vary significantly.

\section{Discussion}
\label{sec:discussion}

The perturbation measurements for this cluster sample constrain the average Mach number of the systems. The sample covers a wide range of dynamic ICM-states providing insight to the influence of different perturbation events.

\subsection{Caveats of perturbation measurements}
\label{sec:sam_turbcav}

It has been shown by \citet{2014MNRAS.444.1497S} that in the PKS0745-191 cluster projection effects in the measured parameters caused the calculated projected spread to be only about half of the real spread in the ICM. Therefore we expect projection effects to strongly affect the absolute values of our spread parameters. But since the effect seems to be similar on all parameters in the PKS0745-191 system the comparison of ratios between different spreads should not be affected. 
All asymmetry measurements are based on circular extraction regions, which means ellipticity of the cluster emission, as in PKS0745-191, will add to the spread values. Thus strong ellipticity could also influence the measured ratios between perturbations.
\citet{2014MNRAS.444.1497S} showed for PKS0745-191 that different bin sizes within a certain range lead to consistent results in the spread analysis. However if the bins are too large like for some more distant clusters in our sample, we generally obtain larger absolute spread-values (e.g. zw3146 in Fig. \ref{fig:dsdp}). Overall the comparison of spread measurements on maps with a S/N of 25 and 50 were in good agreement.

In real clusters there are many factors influencing perturbation measurements. The clusters that lie outside the relations expected from simulations seem to be dominated by processes that have not been included in the simulations (like mergers, AGN feedback bubbles, shocks, or uplifted cold gas). Also projection effects in our measurements and the overall geometry of a system could cause deviations.
Note that the parameters dS and dP are not independent, since they are derived from the independent fit parameters T and $\rm{\eta}$ (see Sect. \ref{sec:maps}).
Uncertainties in the spectral fits are larger at higher temperatures. The size of the error bars of individual measurements limit the sensitivity for finding additional spread in the MCMC calculations (see Sect. \ref{sec:asym}). 
In the case of Abell\,1689 the absolute spread measurements are very low and we only obtain upper limits for this system, which is caused by very low fluctuations and high temperatures, lowering the sensitivity for detecting additional spread.

With the caveats described above our measurements allow for a rough estimate of the average Mach number and thus turbulence trends in the ICM for a large sample of massive clusters of galaxies. Future simulations will help to better quantify the influence of the above described caveats on the Mach number estimates.

\subsection{Relating perturbations to turbulence}
\label{sec:sam_turbturb}

\begin{figure}
  \resizebox{\hsize}{!}{\includegraphics[angle=0,clip=]{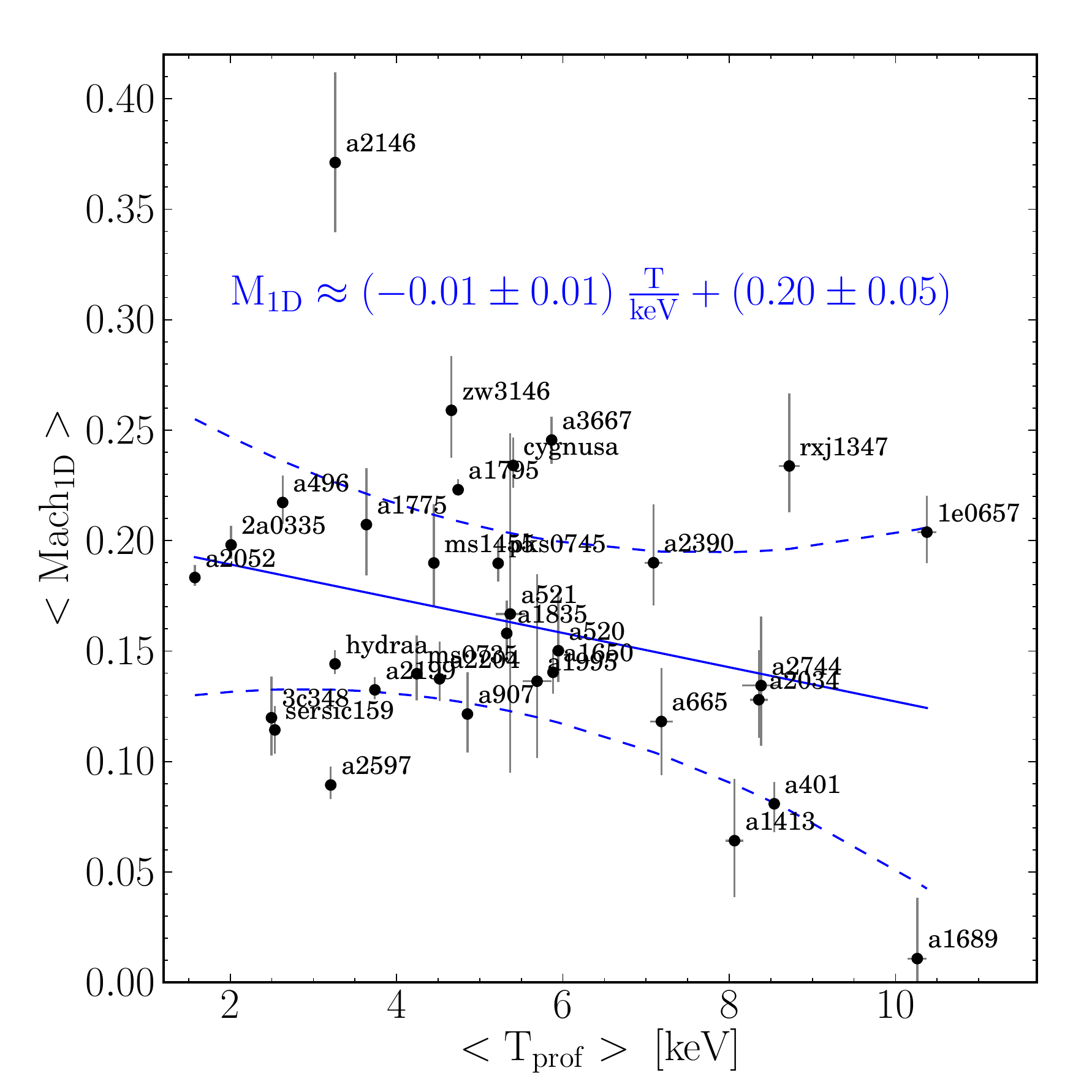}}
  \caption{
    Relation between the average cluster temperature $\rm{<T~prof>}$ and the fractional spread value of the dominating perturbation (dS or dP), which is proportional to the 1D Mach number. The solid and dashed lines represent the best linear correlation and its $\rm{1\sigma}$ scatter. Error bars are the statistical uncertainty from the MCMC measurements.}
  \label{fig:dst}
\end{figure}

Recent high-resolution 3D hydrodynamic simulations by \citet{2014A&A...569A..67G} show that entropy and pressure drive the main perturbations in the ICM depending on the Mach number in the medium. In the low Mach number case ($\rm{Mach_{3D} < 0.5}$) perturbations are mainly isobaric, implying:\\
\\
dP/P is negligible, $\rm{dS/S\sim Mach_{1D}}$, and $\rm{dn/n\sim \mid dT/T \mid}$\\
\\
For large Mach numbers ($\rm{Mach_{3D} > 0.5}$) perturbations shift to the adiabatic regime (because turbulence becomes more violent, overcoming the cluster stratification), implying:\\
\\
dS/S becomes negligible, $\rm{dP/P\sim Mach_{1D}}$, \\and $\rm{dn/n\sim 1/(\gamma -1)~dT/T\sim 1.5~dT/T}$\\
\\
The measurements of the average perturbations in the sample suggest a mixture of the two states (see Figs. \ref{fig:dsdp} and \ref{fig:dtdn}). We observed that entropy perturbations are slightly dominating pressure fluctuations (Fig. \ref{fig:dsdp}) and density fluctuations are comparable to temperature fluctuations with a possible tendency of slightly dominating dn (Fig. \ref{fig:dtdn}). 
Enhanced temperature asymmetry can be caused by displaced cold gas from the central region due to strong feedback like e.g. in Cygnus\,A or strong cold fronts as observed in Abell\,3667. Additional conduction in the ICM can weaken temperature perturbations \citep[][]{2014A&A...569A..67G}.

Assuming entropy perturbations dominate in the sample would it follows that $\rm{dS/S\sim Mach_{1D}}$. The dS/S value is related to the normalisation of the power spectrum, which is related to the peak of the power spectrum, that occurs at large physical radii (low wavenumber k), typically $\rm{\gtrsim100~kpc}$ or $\rm{\gtrsim0.1~r_{500}}$. As an approximation we used the overall average of perturbations (M=1, see Sect. \ref{sec:asym}) as turbulence indicator since the assumption that dS dominates does not hold in the outer regions $\rm{\gtrsim100~kpc}$ of the clusters (see Sect. \ref{sec:sam_inout}). The measurements on average suggest $\rm{Mach_{1D}\approx0.16\pm0.07}$ (see Fig. \ref{fig:dst}).\\
Hydrodynamical simulations of galaxy clusters usually find the ratio between turbulence energy $\rm{E_{turb}}$ and the thermal energy $\rm{E_{therm}}$ to be $\rm{E_{turb}\approx3-30\%~E_{therm}}$, from relaxed to unrelaxed clusters \citep[][]{1999LNP...530..106N, 2009A&A...504...33V, 2009ApJ...705.1129L, 2011A&A...529A..17V, 2012A&A...544A.103V, 2012ApJ...746...94G, 2014ApJ...782...21M, 2014MNRAS.440.3051S}. Since,

\begin{equation}
 \rm{E_{turb}=0.5~\gamma~(\gamma-1)~Mach_{3D}^2~E_{therm}\simeq0.56~Mach_{3D}^2~E_{therm}}
\end{equation}

where the adiabatic index $\rm{\gamma=C_P/C_V=5/3}$ ($\rm{C_P, C_V}$ heat capacity of the medium at constant pressure and volume respectively), it follows that $\rm{Mach_{3D}\simeq0.23-0.73}$ and thereby,

\begin{equation}
 \rm{Mach_{1D}=Mach_{3D}/\sqrt{3} \simeq 0.13-0.42}
\end{equation}

which is consistent with the measured average of $\rm{Mach_{1D}\approx0.16\pm0.07}$. This would suggest that turbulence energy in this sample of clusters is on average about four per cent of the thermal energy in the systems.
Fig. \ref{fig:dst} indicates a weak anti-correlation with average cluster temperature. In the best fit linear regression $\rm{Mach_{1D}\approx-(0.01\pm0.01)~<T_{prof}>/keV + (0.20\pm0.05)}$, the slope is consistent with zero within the errors. 
The significance of the slope was estimated using the re-sampling technique of Sect. \ref{sec:sam_met} by investigating the distribution of fit functions in 1000 fits to random sub-samples of half the size of the original.
Note that the absolute values used as Mach number indicator are subject to some uncertainty like projection- and feedback-effects that have not been taken into account in the simulations we compare to (see Sect. \ref{sec:sam_turbcav}).

\subsection{Difference between core and outskirts}
\label{sec:diss_inout}

If the enhanced pressure perturbations in the outer regions of the clusters in this sample can be confirmed this could be interpreted as a change in the thermodynamic state of the ICM from the center to the outer regions. With increasing pressure perturbations dP we expect more turbulent, pressure-wave driven, motions in the ICM.

This fits the expectation of less relaxed outer regions in clusters where the ICM is not yet virialised and still being accreted into the cluster potential \citep[e.g.][]{2009ApJ...705.1129L}. The change of the dP/dS ratio could also be due to the dependence of the ratio on the probed scales \citep[e.g.][]{2013A&A...559A..78G,2014A&A...569A..67G}. In annuli at larger radii we probe larger scales of ICM fluctuations. A change in dP/dS with scale would also show a radial dependence in the sample.

\subsection{ICM metallicity}
\label{sec:diss_met}

The metallicity of the ICM and its local distribution are of great interest when studying the processes which are enriching the ICM with heavier nuclei. They are thought to be mainly produced by supernova explosions in the galaxies (mainly in the brightest clusters galaxy, BCG) within the clusters and then transported into the ICM \citep[see e.g.][]{2010A&ARv..18..127B,2007A&A...465..345D}.

The observed anti-correlation between average cluster temperature and metallicity (Fig. \ref{fig:metall}) could have many different causes. \citet{2004MNRAS.349..952S} found a similar relation between metallicity and temperature in the Perseus cluster. After testing for various systematic effects they found the correlation to be real but no definite explanation for the effect has been found so far.
Sloshing structures of colder gas have been found to coincide with higher metallicity \citep[see e.g.][]{2011MNRAS.413.2057R}, which could partially explain an anti-correlation.

The temperatures of the clusters in our sample are correlated to their redshift and the metallicity anti-correlates with redshift as well as temperature. \citet{2007A&A...462..429B} probed a sample of clusters in the redshift range $\rm{0.3 < z < 1.3}$ and found significant evolution in metallicity between their higher and lower redshift clusters by more than a factor of two. This trend was confirmed by \citet{2008ApJS..174..117M}. Both studies also found a correlation between cluster temperature and metallicity. However e.g. \citet{2012A&A...537A.142B} found no significant evolution of abundance with redshift. It is not clear whether to expect any evolution within the narrow redshift range ($\rm{0.025 \leq z \leq 0.45}$) we probed. This corresponds to a time span of $\rm{\Delta t \approx 4~Gyr}$ in the standard cosmology assumed in this study.

In individual clusters there is a general trend of lower metallicity at larger radii \citep[see also][]{2001ApJ...551..153D,2004MNRAS.349..952S,2008A&A...487..461L} which means the average metallicity will be lower when we cover larger radial ranges for clusters at higher redshift. The metallicity drops from the centres ($\rm{Z_{solar}\approx0.4-0.8}$) with typically lower temperature gas out to $\rm{\sim300~kpc}$ where the profile flattens around 0.2 solar metallicity fraction. Uniform metallicity distribution at larger radii has been observed in great detail in the Perseus cluster \citep[][]{2013Natur.502..656W}.

The colder gas generally resides in the cluster centres, often in the vicinity of large cD galaxies, that might cause enhanced enrichment. More massive (higher temperature) systems will also have undergone more mergers and generally have stronger AGN feedback, which contributes to the dilution of metals in the X-ray halo \citep[e.g.][]{2011MNRAS.411..349G} and affects the average area- and error-weighted metallicity calculated in this study. Also multiphasedness of the hot ICM along the line of sight could influence the correlation as we typically probe larger volumes for higher temperatures.
Multiphase gas has been found to bias the ICM metallicity measurements from X-ray spectra in some clusters \citep[e.g.][]{2013MNRAS.433.3290P}. Unresolved 2D structure of the ICM could bias the measured metallicity to higher values for systems with intermediate temperatures \citep[see][]{2008ApJ...674..728R,2009A&A...493..409S,2010A&A...522A..34G}. This study reduced multiphase effects by resolving the spatial structure and measuring the average metallicity of many spatial-spectral bins.

Another factor of influence could be that star formation efficiency decreases with cluster mass and temperature \citep[e.g.][]{2010A&ARv..18..127B}. There have been many studies in the infrared wavelength \citep[e.g.][]{2012A&A...537A..58P,2015A&A...574A.105P} which show quenching of star formation in massive DM halos of galaxy clusters and groups. Our sample predominantly consists of relatively massive systems (see Tab. \ref{tab:sampro}) at low to intermediate redshifts and in different stages of evolution. Note that all metallicity measurements are based on a fixed solar abundance model \citep[][]{1989GeCoA..53..197A}.

\section{Summary and conclusions}
\label{sec:conclusion}

We presented a very large sample of detailed cluster maps with application to understand the thermodynamic processes in clusters of galaxies. The deep observations of the individual clusters helped to identify structures in the ICM caused by mergers or AGN feedback. By comparing to recent high-resolution simulations of perturbations in the ICM we constrained the average 1D Mach number regime in the sample to $\rm{Mach_{1D}\approx0.16\pm0.07}$ with some caveats (see Sect. \ref{sec:discussion}). Comparing to simulations this would suggest $\rm{E_{turb}\approx0.04~E_{therm}}$ (see Sect. \ref{sec:sam_turbturb}). By comparing perturbations in the central regions ($\rm{\lesssim100~kpc}$) and in the outer regions ($\rm{\gtrsim100~kpc}$) we found an indication for a change in the thermodynamic state from mainly isobaric to a more adiabatic regime. 

In addition the sample shows a tight correlation between the average cluster metallicity, average temperature and redshift. The best fit linear correlation between metallicity and redshift is $\rm{Z/Z_\odot=-(0.6\pm0.2)~z+(0.36\pm0.04)}$ and between metallicity and temperature the best fit is $\rm{Z/Z_\odot = -(1.0\pm0.7)~T/100keV + (0.34\pm0.06)}$.
The average metallicity of the sample is $\rm{Z\approx0.3\pm0.1~Z_\odot}$. 

Future X-ray missions like Astro-H \citep[][]{2014arXiv1412.1176K} and Athena \citep[][]{2013arXiv1306.2307N} will help to further investigate turbulent velocities and chemical enrichment in the ICM. The eROSITA observatory \citep[][]{2012arXiv1209.3114M} will detect a large X-ray cluster sample for cosmological studies and our detailed cluster mapping can be used to make predictions on the scatter in scaling relations due to unresolved structures in temperature. To encourage further analysis based on this unique sample of cluster observations all maps and asymmetry measurements used in this study are made publicly available in electronic form.

\begin{acknowledgements}

We thank the anonymous referee for constructive comments that helped to improve the clarity of the paper. 
We thank J. Buchner for helpful discussions.
This research has made use of data obtained from the \emph{Chandra} Data Archive and the \emph{Chandra} Source Catalog, and software provided by the \emph{Chandra} X-ray Center (CXC) in the application packages CIAO, ChIPS, and Sherpa. 
This research has made use of NASA's Astrophysics Data System. 
This research has made use of the VizieR catalogue access tool, CDS, Strasbourg, France. 
This research has made use of SAOImage DS9, developed by Smithsonian Astrophysical Observatory. 
This research has made use of data and software provided by the High Energy Astrophysics Science Archive Research Center (HEASARC), which is a service of the Astrophysics Science Division at NASA/GSFC and the High Energy Astrophysics Division of the Smithsonian Astrophysical Observatory. 
This research has made use of the SIMBAD database, operated at CDS, Strasbourg, France. 
This research has made use of the tools Veusz, the matplotlib library for Python, and TOPCAT.
M.G. is supported by the National Aeronautics and Space Administration through Einstein Postdoctoral Fellowship Award Number PF-160137 issued by the \emph{Chandra} X-ray Observatory Center, which is operated by the Smithsonian Astrophysical Observatory for and on behalf of the National Aeronautics Space Administration under contract NAS8-03060.

\end{acknowledgements}

\bibliographystyle{aa}
\bibliography{auto,general}

\Online
\begin{appendix}

\section{Cluster sample}
\label{sec:aobs}
\begin{center}
%--------------------------
\begin{table*}[h!]
\caption[]{\emph{Chandra} cluster sample (CIZA clusters).}
\begin{center}
\begin{tabular}{lrrrrrlr}
\hline\hline\noalign{\smallskip}
  \multicolumn{1}{l}{Cluster\tablefootmark{a}} &
  \multicolumn{1}{l}{Exp\tablefootmark{b}} &
  \multicolumn{1}{l}{RA (J2000)} &
  \multicolumn{1}{l}{DEC (J2000)} &
  \multicolumn{1}{l}{Flux\tablefootmark{c}} &
  \multicolumn{1}{l}{z} &
  \multicolumn{1}{l}{Obj. ID (CIZA)} &
  \multicolumn{1}{l}{$\rm{L_X}$\tablefootmark{d}} \\
  \multicolumn{1}{l}{} &
  \multicolumn{1}{l}{[ks]} &
  \multicolumn{1}{l}{[deg]} &
  \multicolumn{1}{l}{[deg]} &
  \multicolumn{1}{l}{[$\rm{10^{-15}~W/m^2}$]} &
  \multicolumn{1}{l}{} &
  \multicolumn{1}{l}{} &
  \multicolumn{1}{l}{[$\rm{10^{37}~W}$]} \\
\noalign{\smallskip}\hline\noalign{\smallskip}
  CYGNUS A & 232 & 299.877 & 40.741 & 52.82 & 0.0561 & J1959.5+4044 & 7.08\\
  PKS 0745-191 & 174 & 116.883 & -19.290 & 45.73 & 0.1028 & J0747.5-1917 & 20.36\\
\noalign{\smallskip}\hline
\end{tabular}
\tablefoot{
\tablefoottext{a}{Most commonly used name of cluster or central object.}
\tablefoottext{b}{Combined ACIS-S/-I exposure after excluding times of high background.}
\tablefoottext{c}{ROSAT 0.1-2.4\,keV X-ray flux in $\rm{10^{-15}~W/m^2}$}
\tablefoottext{d}{ROSAT 0.1-2.4\,keV X-ray luminosity in $\rm{10^{37}~W}$ (CIZA column: LX)}
}
\end{center}
\label{tab:cat1}
\end{table*}
%--------------------------
%--------------------------
\begin{table*}[h!]
\caption[]{\emph{Chandra} cluster sample (NORAS clusters).}
\begin{center}
\begin{tabular}{lrrrrrlr}
\hline\hline\noalign{\smallskip}
  \multicolumn{1}{l}{Cluster\tablefootmark{a}} &
  \multicolumn{1}{l}{Exp\tablefootmark{b}} &
  \multicolumn{1}{l}{RA (J2000)} &
  \multicolumn{1}{l}{DEC (J2000)} &
  \multicolumn{1}{l}{Flux\tablefootmark{c}} &
  \multicolumn{1}{l}{z} &
  \multicolumn{1}{l}{Obj. ID (NORAS)} &
  \multicolumn{1}{l}{$\rm{L_X}$\tablefootmark{d}} \\
  \multicolumn{1}{l}{} &
  \multicolumn{1}{l}{[ks]} &
  \multicolumn{1}{l}{[deg]} &
  \multicolumn{1}{l}{[deg]} &
  \multicolumn{1}{l}{[$\rm{10^{-15}~W/m^2}$]} &
  \multicolumn{1}{l}{} &
  \multicolumn{1}{l}{} &
  \multicolumn{1}{l}{[$\rm{10^{37}~W}$]} \\
\noalign{\smallskip}\hline\noalign{\smallskip}
  A 2052 & 651 & 229.182 & 7.013 & 47.94 & 0.0353 & RXC J1516.7+0701 & 2.58\\
  A 1775 & 99 & 205.448 & 26.352 & 12.58 & 0.0724 & RXC J1341.8+2622 & 2.83\\
  A 2199 & 156 & 247.188 & 39.553 & 97.92 & 0.0299 & RXC J1628.6+3932 & 3.77\\
  2A 0335+096 & 102 & 54.665 & 10.007 & 80.91 & 0.0347 & RXC J0338.6+0958 & 4.21\\
  3C348 (HERCULES A) & 112 & 252.778 & 4.985 & 5.39 & 0.154 & RXC J1651.1+0459 & 5.49\\
  A 2034 & 255 & 227.532 & 33.515 & 11.94 & 0.113 & RXC J1510.1+3330 & 6.49\\
  MS0735.6+7421 & 520 & 115.421 & 74.266 & 4.06 & 0.2149 & RXC J0741.7+7414 & 7.94\\
  A 2146 & 418 & 239.006 & 66.352 & 3.99 & 0.2339 & RXC J1556.1+6621 & 9.31\\
  A 1795 & 958 & 207.221 & 26.596 & 59.29 & 0.0622 & RXC J1348.8+2635 & 9.93\\
  A 1413 & 136 & 178.769 & 23.369 & 12.61 & 0.1427 & RXC J1155.3+2324 & 10.91\\
  A 401 & 163 & 44.727 & 13.579 & 50.29 & 0.0739 & RXC J0258.9+1334 & 11.76\\
  A 1995 & 100 & 223.168 & 58.049 & 3.18 & 0.3179 & RXC J1452.9+5802 & 13.42\\
  MS 1455.0+2232 & 108 & 224.253 & 22.33 & 4.89 & 0.2579 & RXC J1457.2+2220 & 13.73\\
  A 520 & 527 & 73.546 & 2.977 & 8.33 & 0.203 & RXC J0454.1+0255 & 14.52\\
  A 665 & 140 & 127.637 & 65.89 & 11.18 & 0.1818 & RXC J0830.9+6551 & 15.69\\
  A 2204 & 97 & 248.186 & 5.557 & 24.11 & 0.1514 & RXC J1632.7+0534 & 23.43\\
  A 2390 & 110 & 328.403 & 17.683 & 11.01 & 0.2329 & RXC J2153.5+1741 & 25.15\\
  ZW 3146 & 84 & 155.906 & 4.167 & 8.77 & 0.285 & RXC J1023.6+0411 & 29.91\\
  A 1835 & 223 & 210.271 & 2.895 & 12.12 & 0.2528 & RXC J1401.0+0252 & 32.56\\
\noalign{\smallskip}\hline
\end{tabular}
\tablefoot{
\tablefoottext{a}{Most commonly used name of cluster or central object.}
\tablefoottext{b}{Combined ACIS-S/-I exposure after excluding times of high background.}
\tablefoottext{c}{ROSAT 0.1-2.4\,keV X-ray flux in $\rm{10^{-15}~W/m^2}$}
\tablefoottext{d}{ROSAT 0.1-2.4\,keV X-ray luminosity in $\rm{10^{37}~W}$ (NORAS column: LX)}
}
\end{center}
\label{tab:cat2}
\end{table*}
%--------------------------
%--------------------------
\begin{table*}[h!]
\caption[]{\emph{Chandra} cluster sample (REFLEX clusters).}
\begin{center}
\begin{tabular}{lrrrrrlr}
\hline\hline\noalign{\smallskip}
  \multicolumn{1}{l}{Cluster\tablefootmark{a}} &
  \multicolumn{1}{l}{Exp\tablefootmark{b}} &
  \multicolumn{1}{l}{RA (J2000)} &
  \multicolumn{1}{l}{DEC (J2000)} &
  \multicolumn{1}{l}{Flux\tablefootmark{c}} &
  \multicolumn{1}{l}{z} &
  \multicolumn{1}{l}{Obj. ID (REFLEX)} &
  \multicolumn{1}{l}{$\rm{L_X}$\tablefootmark{d}} \\
  \multicolumn{1}{l}{} &
  \multicolumn{1}{l}{[ks]} &
  \multicolumn{1}{l}{[deg]} &
  \multicolumn{1}{l}{[deg]} &
  \multicolumn{1}{l}{[$\rm{10^{-15}~W/m^2}$]} &
  \multicolumn{1}{l}{} &
  \multicolumn{1}{l}{} &
  \multicolumn{1}{l}{[$\rm{10^{37}~W}$]} \\
\noalign{\smallskip}\hline\noalign{\smallskip}
  SERSIC 159-03 & 106 & 348.515 & -42.713 & 23.412 & 0.0564 & J2313.9-4244 & 3.74\\
  A 496 & 88 & 68.403 & -13.25 & 72.075 & 0.0326 & J0433.6-1315 & 3.89\\
  HYDRA A & 224 & 139.527 & -12.092 & 39.461 & 0.0539 & J0918.1-1205 & 5.61\\
  A 1650 & 229 & 194.664 & -1.781 & 20.909 & 0.0845 & J1258.6-0145 & 6.99\\
  A 2597 & 146 & 351.337 & -12.136 & 20.558 & 0.0852 & J2325.3-1207 & 7.22\\
  A 3667 & 528 & 303.211 & -56.855 & 70.892 & 0.0556 & J2012.5-5649 & 10.02\\
  A 907 & 103 & 149.528 & -11.086 & 7.833 & 0.1669 & J0958.3-1103 & 10.13\\
  A 521 & 165 & 73.558 & -10.273 & 4.944 & 0.2475 & J0454.1-1014 & 12.97\\
  A 2744 & 124 & 3.586 & -30.352 & 4.964 & 0.3066 & J0014.3-3023 & 19.79\\
  A 1689 & 197 & 197.808 & -1.337 & 15.332 & 0.1832 & J1311.4-0120 & 23.59\\
  1E 0657-56 & 566 & 104.751 & -55.904 & 9.079 & 0.2965 & J0658.5-5556 & 35.55\\
  RX J1347-114 & 232 & 206.889 & -11.734 & 6.468 & 0.4516 & J1347.5-1144 & 63.43\\
\noalign{\smallskip}\hline
\end{tabular}
\tablefoot{
\tablefoottext{a}{Most commonly used name of cluster or central object.}
\tablefoottext{b}{Combined ACIS-S/-I exposure after excluding times of high background.}
\tablefoottext{c}{ROSAT 0.1-2.4\,keV X-ray flux in $\rm{10^{-15}~W/m^2}$}
\tablefoottext{d}{ROSAT 0.1-2.4\,keV X-ray luminosity in $\rm{10^{37}~W}$ (REFLEX column: LumCor, h=0.5)}
}
\end{center}
\label{tab:cat3}
\end{table*}
%--------------------------
\end{center}

\section{\emph{Chandra} datasets}
\label{sec:cobs}

%--------------------------
\begin{table*}[h!]
\caption[]{\emph{Chandra} datasets used in this study.}
\begin{center}
\begin{tabular}{ll}
\hline\hline\noalign{\smallskip}
  \multicolumn{1}{l}{Cluster\tablefootmark{a}} &
  \multicolumn{1}{l}{\emph{Chandra} ObsID\tablefootmark{b}} \\
\noalign{\smallskip}\hline\noalign{\smallskip}

1e0657 & 554,3184,4984,4985,4986,5355,5356,5357,5358,5361 \\  
2a0335 & 919,7939,9792 \\  
3c348 & 1625,5796,6257 \\  
a1413 & 537,1661,5002,5003,7696 \\  
a1650 & 4178,5822,5823,6356,6357,6358,7242,7691 \\  
a1689 & 540,1663,5004,6930,7289,7701 \\  
a1775 & 12891,13510 \\  
a1795 & 493,494,3666,5286,5287,5288,5289,5290,6159,6160,6161,6162,6163,\\
&10432,10433,10898,10899,10900,10901,12026,12027,12028,12029,13106,13107,\\
&13108,13109,13110,13111,13112,13113,13412,13413,13414,13415,13416,13417,\\
&14268,14269,14270,14271,14272,14273,14274,14275,15485,15486,15487,15488,15489,\\
&15490,15491,15492,16432,16433,16434,16435,16436,16437,16438,16439,16465,16466,\\
&16467,16468,16469,16470,16471,16472 \\  
a1835 & 495,496,6880,6881,7370 \\  
a1995 & 906,7021,7713 \\  
a2034 & 2204,7695,12885,12886,13192,13193 \\  
a2052 & 890,5807,10477,10478,10479,10480,10879,10914,10915,10916,10917 \\  
a2146 & 10464,10888,12245,12246,12247,13020,13021,13023,13120,13138 \\  
a2199 & 497,498,10748,10803,10804,10805 \\  
a2204 & 499,6104,7940 \\  
a2390 & 500,501,4193 \\  
a2597 & 922,6934,7329,15144 \\  
a2744 & 2212,7712,7915,8477,8557 \\  
a3667 & 513,889,5751,5752,5753,6292,6295,6296,7686 \\  
a401 & 518,2309,14024 \\  
a496 & 931,3361,4976 \\  
a520 & 528,4215,7703,9424,9425,9426,9430 \\  
a521 & 430,901,12880,13190 \\  
a665 & 531,3586,7700,12286,13201,15148 \\  
a907 & 535,3185,3205 \\  
cygnusa & 360,5830,5831,6225,6226,6228,6229,6250,6252 \\  
hydraa & 575,576,4969,4970 \\  
ms0735 & 4197,10468,10469,10470,10471,10822,10918,10922 \\  
ms1455 & 543,4192,7709 \\  
pks0745 & 508,2427,6103,7694,12881 \\  
rxj1347 & 506,507,3592,13516,13999,14407 \\  
sersic159 & 1668,11758 \\  
zw3146 & 909,9371 \\

\noalign{\smallskip}\hline
\end{tabular}
\tablefoot{
\tablefoottext{a}{Abbreviated cluster name.}
\tablefoottext{b}{List of \emph{Chandra} observations used in this study (indicated by their ObsID, observation identification number).}
}
\end{center}
\label{tab:obsids}
\end{table*}
%--------------------------

\section{Perturbation table}
\label{sec:amaps}

%--------------------------
\begin{table*}[h!]
\caption[]{Measured fractional perturbations.}
\begin{center}
\begin{tabular}{lrrrr}
\hline\hline\noalign{\smallskip}
  \multicolumn{1}{l}{Cluster\tablefootmark{a}} &
  \multicolumn{1}{l}{dP\tablefootmark{b}} &
  \multicolumn{1}{l}{dS\tablefootmark{b}} &
  \multicolumn{1}{l}{dT\tablefootmark{b}} &
  \multicolumn{1}{l}{dn\tablefootmark{b}} \\
  \multicolumn{1}{l}{} &
  \multicolumn{1}{l}{[per cent]} &
  \multicolumn{1}{l}{[per cent]} &
  \multicolumn{1}{l}{[per cent]} &
  \multicolumn{1}{l}{[per cent]} \\
\noalign{\smallskip}\hline\noalign{\smallskip}

1e0657 & $\rm{20.4^{+1.6}_{-1.4}}$  & $\rm{17.3^{+2.1}_{-1.5}}$ & $\rm{8.8^{+1.5}_{-1.4}}$ & $\rm{20.9^{+1.2}_{-0.9}}$ \\[2pt]
2a0335 & $\rm{11.7^{+0.6}_{-0.5}}$  & $\rm{19.8^{+0.9}_{-0.6}}$ & $\rm{9.3^{+0.5}_{-0.4}}$ & $\rm{13.5^{+0.6}_{-0.4}}$ \\[2pt]
3c348 & $\rm{1.3^{+2.9}_{-1.3}}$  & $\rm{12.0^{+1.9}_{-1.7}}$ & $\rm{3.6^{+2.0}_{-2.8}}$ & $\rm{8.5^{+0.9}_{-0.6}}$ \\[2pt]
a1413 & $\rm{5.4^{+1.9}_{-4.0}}$  & $\rm{6.4^{+2.8}_{-2.6}}$ & $\rm{0.3^{+2.4}_{-0.3}}$ & $\rm{8.7^{+1.0}_{-0.7}}$ \\[2pt]
a1650 & $\rm{5.5^{+0.9}_{-1.0}}$  & $\rm{14.0^{+1.0}_{-1.0}}$ & $\rm{4.0^{+1.0}_{-1.0}}$ & $\rm{10.0^{+0.7}_{-0.5}}$ \\[2pt]
a1689 & $\rm{0.6^{+2.7}_{-0.6}}$  & $\rm{1.1^{+2.8}_{-1.1}}$ & $\rm{0.8^{+2.1}_{-0.8}}$ & $\rm{3.8^{+0.4}_{-0.2}}$ \\[2pt]
a1775 & $\rm{0.6^{+3.5}_{-0.6}}$  & $\rm{20.7^{+2.6}_{-2.3}}$ & $\rm{12.6^{+1.9}_{-2.1}}$ & $\rm{8.0^{+0.8}_{-0.8}}$ \\[2pt]
a1795 & $\rm{5.3^{+0.2}_{-0.2}}$  & $\rm{22.3^{+0.5}_{-0.2}}$ & $\rm{11.7^{+0.4}_{-0.1}}$ & $\rm{12.4^{+0.3}_{-0.1}}$ \\[2pt]
a1835 & $\rm{6.0^{+1.1}_{-1.1}}$  & $\rm{15.8^{+1.5}_{-1.1}}$ & $\rm{9.3^{+1.0}_{-0.8}}$ & $\rm{8.8^{+0.6}_{-0.5}}$ \\[2pt]
a1995 & $\rm{5.7^{+5.5}_{-5.7}}$  & $\rm{13.6^{+4.9}_{-3.5}}$ & $\rm{11.4^{+3.7}_{-5.8}}$ & $\rm{7.0^{+1.6}_{-0.9}}$ \\[2pt]
a2034 & $\rm{12.8^{+2.2}_{-1.7}}$  & $\rm{6.5^{+2.1}_{-2.4}}$ & $\rm{6.1^{+1.6}_{-3.4}}$ & $\rm{9.1^{+1.0}_{-0.7}}$ \\[2pt]
a2052 & $\rm{7.5^{+0.2}_{-0.2}}$  & $\rm{18.3^{+0.6}_{-0.4}}$ & $\rm{11.6^{+0.4}_{-0.2}}$ & $\rm{10.1^{+0.3}_{-0.2}}$ \\[2pt]
a2146 & $\rm{20.3^{+1.9}_{-1.9}}$  & $\rm{37.1^{+4.1}_{-3.2}}$ & $\rm{12.2^{+1.7}_{-1.4}}$ & $\rm{24.3^{+2.7}_{-1.5}}$ \\[2pt]
a2199 & $\rm{7.6^{+0.4}_{-0.3}}$  & $\rm{13.2^{+0.6}_{-0.4}}$ & $\rm{7.3^{+0.4}_{-0.3}}$ & $\rm{8.6^{+0.3}_{-0.3}}$ \\[2pt]
a2204 & $\rm{6.7^{+1.2}_{-1.3}}$  & $\rm{13.7^{+1.7}_{-1.0}}$ & $\rm{8.7^{+1.1}_{-1.0}}$ & $\rm{7.6^{+0.5}_{-0.5}}$ \\[2pt]
a2390 & $\rm{11.4^{+2.0}_{-1.8}}$  & $\rm{19.0^{+2.6}_{-1.9}}$ & $\rm{12.4^{+2.1}_{-2.0}}$ & $\rm{11.2^{+1.1}_{-0.8}}$ \\[2pt]
a2597 & $\rm{8.1^{+0.8}_{-0.6}}$  & $\rm{8.9^{+0.8}_{-0.6}}$ & $\rm{5.5^{+0.6}_{-0.6}}$ & $\rm{7.2^{+0.5}_{-0.4}}$ \\[2pt]
a2744 & $\rm{9.8^{+4.2}_{-9.8}}$  & $\rm{13.4^{+3.1}_{-2.7}}$ & $\rm{7.6^{+3.5}_{-7.6}}$ & $\rm{15.1^{+1.4}_{-1.1}}$ \\[2pt]
a3667 & $\rm{24.6^{+1.1}_{-1.1}}$  & $\rm{12.4^{+0.7}_{-0.5}}$ & $\rm{16.1^{+0.9}_{-0.7}}$ & $\rm{11.4^{+0.5}_{-0.4}}$ \\[2pt]
a401 & $\rm{6.1^{+1.3}_{-1.5}}$  & $\rm{8.1^{+1.0}_{-1.3}}$ & $\rm{2.8^{+1.3}_{-2.8}}$ & $\rm{7.5^{+0.5}_{-0.3}}$ \\[2pt]
a496 & $\rm{9.0^{+0.6}_{-0.6}}$  & $\rm{21.7^{+1.2}_{-1.0}}$ & $\rm{14.0^{+0.9}_{-0.6}}$ & $\rm{8.9^{+0.5}_{-0.4}}$ \\[2pt]
a520 & $\rm{12.6^{+2.0}_{-1.8}}$  & $\rm{15.0^{+2.6}_{-1.4}}$ & $\rm{7.6^{+1.9}_{-1.7}}$ & $\rm{11.5^{+1.4}_{-0.9}}$ \\[2pt]
a521 & $\rm{12.1^{+5.3}_{-7.8}}$  & $\rm{16.7^{+8.2}_{-7.2}}$ & $\rm{10.0^{+6.4}_{-8.0}}$ & $\rm{10.7^{+2.3}_{-1.4}}$ \\[2pt]
a665 & $\rm{11.8^{+2.4}_{-2.4}}$  & $\rm{8.2^{+2.6}_{-4.6}}$ & $\rm{6.5^{+2.6}_{-3.9}}$ & $\rm{11.0^{+0.9}_{-0.8}}$ \\[2pt]
a907 & $\rm{0.2^{+2.4}_{-0.2}}$  & $\rm{12.1^{+1.9}_{-1.7}}$ & $\rm{1.7^{+2.9}_{-1.7}}$ & $\rm{9.5^{+0.9}_{-0.6}}$ \\[2pt]
cygnusa & $\rm{18.2^{+1.0}_{-0.8}}$  & $\rm{23.4^{+1.3}_{-1.0}}$ & $\rm{14.1^{+0.8}_{-0.6}}$ & $\rm{27.3^{+1.2}_{-0.9}}$ \\[2pt]
hydraa & $\rm{9.8^{+0.4}_{-0.3}}$  & $\rm{14.4^{+0.6}_{-0.5}}$ & $\rm{9.7^{+0.4}_{-0.3}}$ & $\rm{7.8^{+0.3}_{-0.2}}$ \\[2pt]
ms0735 & $\rm{10.6^{+1.5}_{-0.9}}$  & $\rm{14.0^{+1.7}_{-1.2}}$ & $\rm{7.8^{+1.2}_{-0.8}}$ & $\rm{10.3^{+1.0}_{-0.7}}$ \\[2pt]
ms1455 & $\rm{4.1^{+2.0}_{-3.5}}$  & $\rm{19.0^{+2.6}_{-1.9}}$ & $\rm{9.0^{+2.0}_{-1.5}}$ & $\rm{11.1^{+1.0}_{-0.8}}$ \\[2pt]
pks0745 & $\rm{3.4^{+1.0}_{-1.9}}$  & $\rm{19.0^{+1.0}_{-0.8}}$ & $\rm{9.4^{+0.8}_{-0.6}}$ & $\rm{11.6^{+0.4}_{-0.5}}$ \\[2pt]
rxj1347 & $\rm{13.4^{+2.3}_{-2.3}}$  & $\rm{23.4^{+3.3}_{-2.1}}$ & $\rm{16.0^{+2.3}_{-2.0}}$ & $\rm{13.8^{+1.5}_{-0.8}}$ \\[2pt]
sersic159 & $\rm{9.7^{+1.1}_{-0.7}}$  & $\rm{11.4^{+1.1}_{-1.1}}$ & $\rm{5.2^{+0.6}_{-0.6}}$ & $\rm{9.2^{+0.9}_{-0.5}}$ \\[2pt]
zw3146 & $\rm{0.4^{+2.3}_{-0.4}}$  & $\rm{25.9^{+2.5}_{-2.2}}$ & $\rm{10.9^{+1.9}_{-1.9}}$ & $\rm{15.9^{+1.3}_{-0.9}}$ \\[2pt]

\noalign{\smallskip}\hline
\end{tabular}
\tablefoot{
\tablefoottext{a}{Abbreviated cluster name.}
\tablefoottext{b}{Measured average fractional spread of thermodynamic properties for the whole cluster (see Sect. \ref{sec:asym}, M=1) in per cent.}
}
\end{center}
\label{tab:persum}
\end{table*}
%--------------------------

\section{2D Maps}
\label{sec:amaps}

\begin{figure*}[h!]
  \resizebox{\hsize}{!}{\includegraphics[page=1,angle=0,trim=0.3cm 0.3cm 0.3cm 0.3cm,clip=true]{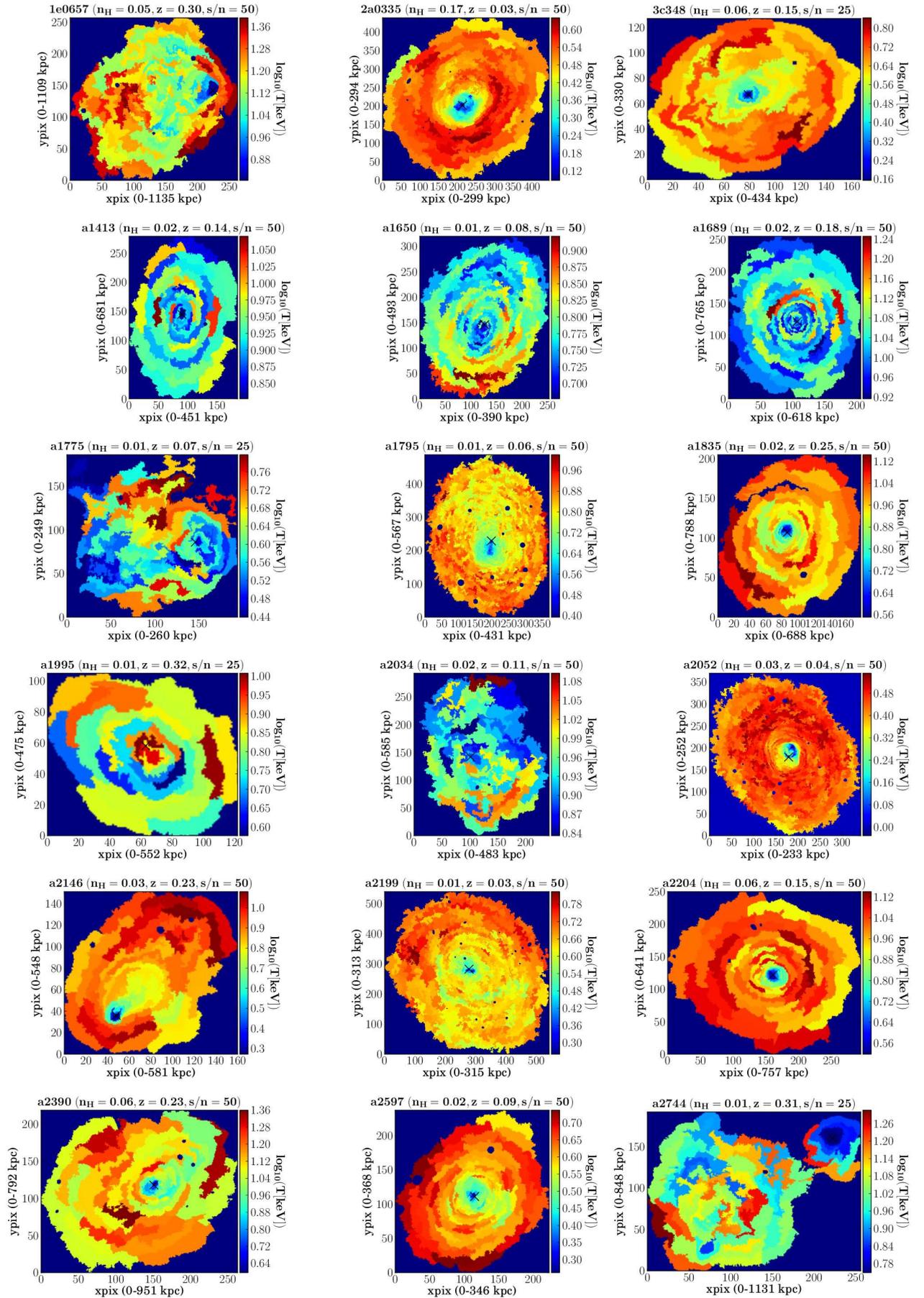}}
  \caption{Temperature maps of all clusters in the sample. Bins have a S/N of 50 or 25 (see Sect. \ref{sec:asym}) and temperatures are shown on a logarithmic colour scale. Excluding areas below the surface brightness cut (area-normalised normalisation $\rm{>10^{-7}~cm^{-5}~arcsec^{-2}}$) and where the errors on temperature are more than twice the best fit temperature value. The plot title gives the abbreviated cluster name, the average foreground column density $\rm{n_H~[cm^{-2}]}$, and the redshift z. Scale: $\rm{1pix\sim1\arcsec}$. Crosses mark the peak of the X-ray emission.}
  \label{fig:Tall}
\end{figure*}
\addtocounter{figure}{-1} 
\begin{figure*}[h!]
  \resizebox{\hsize}{!}{\includegraphics[page=2,angle=0,trim=0.3cm 1.3cm 0.3cm 0.3cm,clip=true]{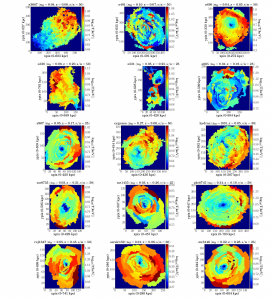}}
  \caption{continued.}
\end{figure*}

\section{Unsharp-masked count images}
\label{sec:amask}

\begin{figure*}[h!]
  \resizebox{\hsize}{!}{\includegraphics[angle=0,trim=0.3cm 0.4cm 0.3cm 0.2cm,clip=true]{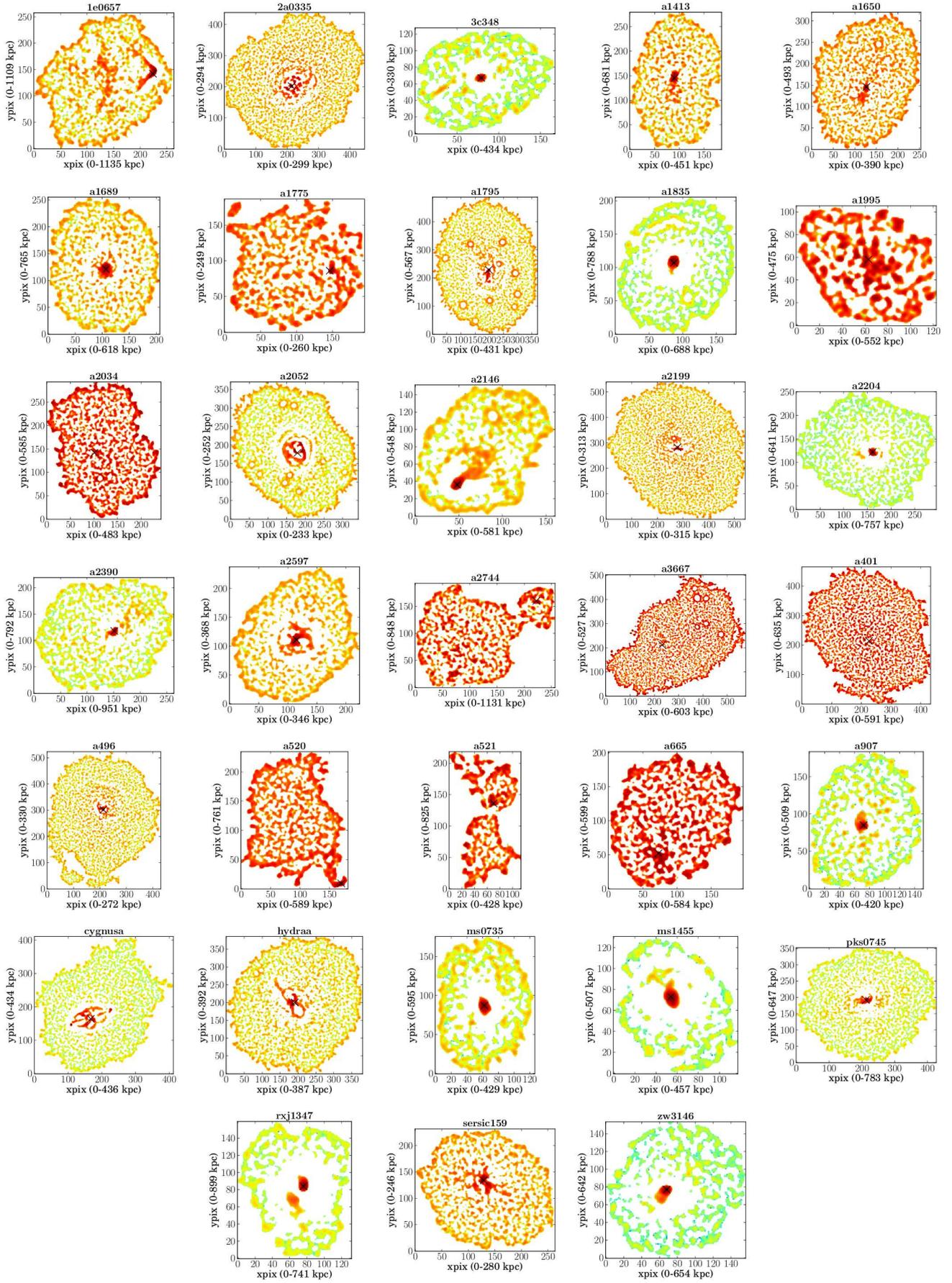}}
  \caption{
    Unsharp-masked exposure-corrected images (logarithmic colour scale) of all clusters in the sample showing the difference between two count images smoothed with a Gaussian function (2 pixels and 5 pixels sigma). Colours indicate relative surface brightness differences (over-densities dark red, under-dense areas green to white). Scale: $\rm{1pix\sim1\arcsec}$. Crosses mark the peak of the X-ray emission.}
  \label{fig:umall}
\end{figure*}

\section{Projected radial profiles}
\label{sec:arad}

\begin{figure*}[h!]
  \resizebox{\hsize}{!}{\includegraphics[page=1,angle=0,trim=0.3cm 0.5cm 0.3cm 0.2cm,clip=true]{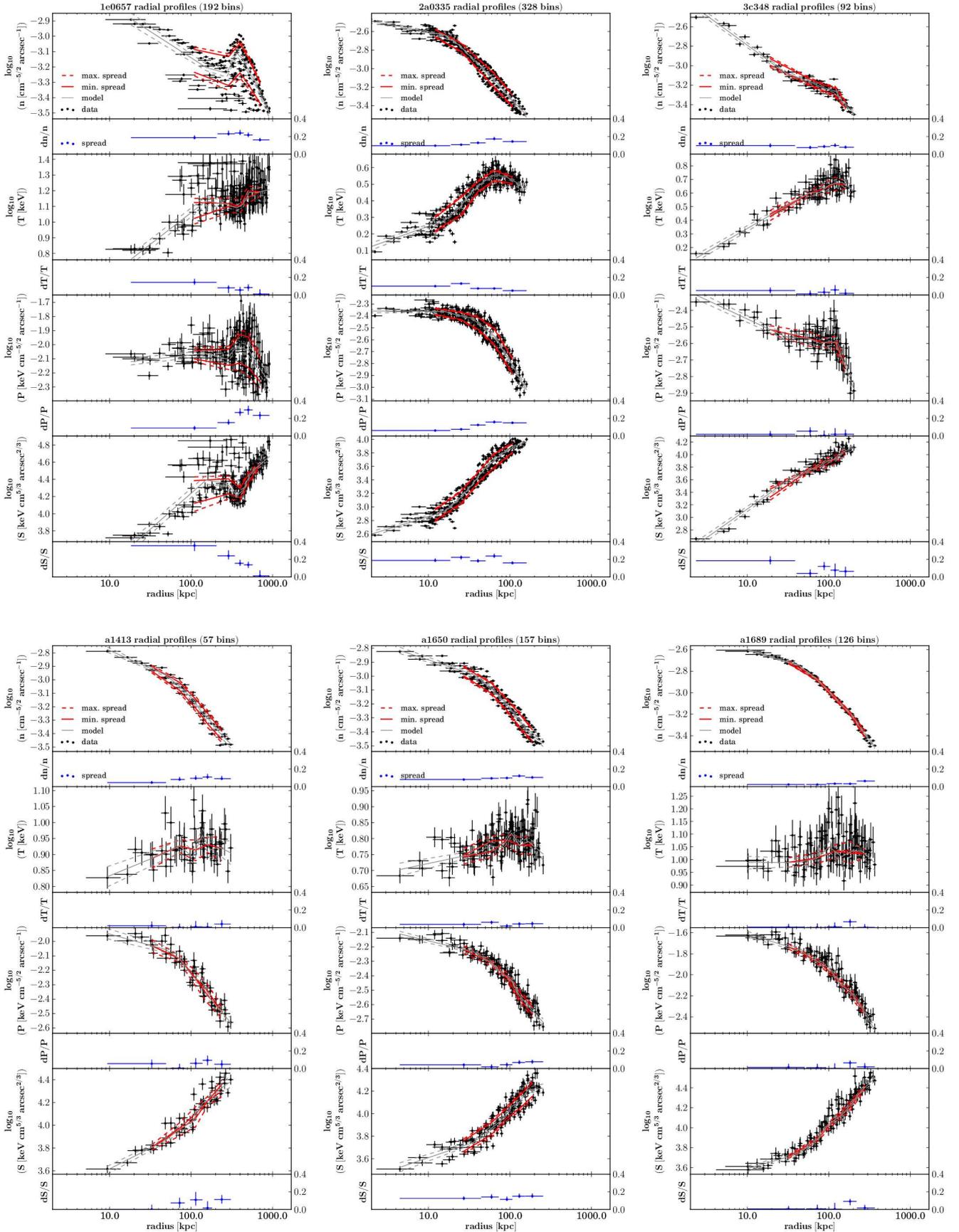}}
  \caption{Radial profiles of projected density, temperature, pressure, and entropy. Cluster names are given in the plot titles. The cluster centres are marked as crosses in Figs. \ref{fig:Tall} and \ref{fig:umall}. Error bars are the fit-errors and the standard deviation of the radial distribution for all spatial-spectral bins. The Plotted lines show limits on intrinsic scatter around an average seven-node model (grey lines) within the given radial range (see Sect. \ref{sec:asym}). The small panels show the measured fractional scatter (M=5) with confidence- and radial-range.}
  \label{fig:Proall}
\end{figure*}
\addtocounter{figure}{-1} 
\begin{figure*}
  \resizebox{\hsize}{!}{\includegraphics[page=2,angle=0,trim=0.3cm 0.5cm 0.3cm 0.2cm,clip=true]{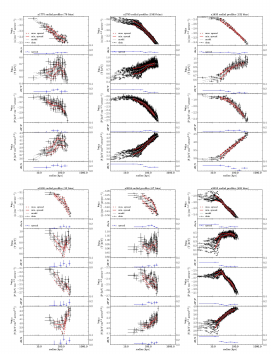}}
  \caption{continued.}
\end{figure*}
\addtocounter{figure}{-1} 
\begin{figure*}
  \resizebox{\hsize}{!}{\includegraphics[page=3,angle=0,trim=0.3cm 0.5cm 0.3cm 0.2cm,clip=true]{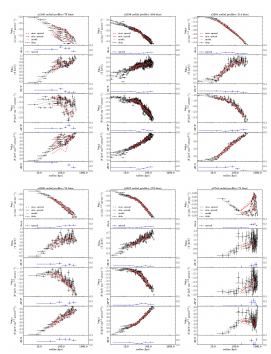}}
  \caption{continued.}
\end{figure*}
\addtocounter{figure}{-1} 
\begin{figure*}
  \resizebox{\hsize}{!}{\includegraphics[page=4,angle=0,trim=0.3cm 0.5cm 0.3cm 0.2cm,clip=true]{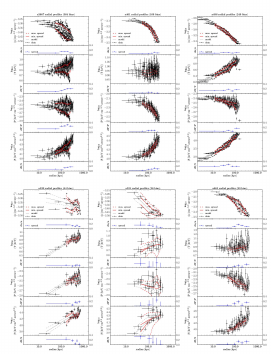}}
  \caption{continued.}
\end{figure*}
\addtocounter{figure}{-1} 
\begin{figure*}
  \resizebox{\hsize}{!}{\includegraphics[page=5,angle=0,trim=0.3cm 0.5cm 0.3cm 0.2cm,clip=true]{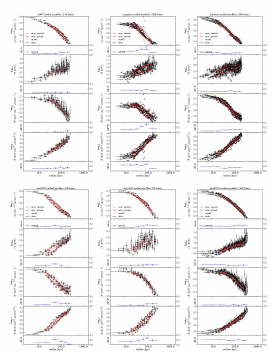}}
  \caption{continued.}
\end{figure*}
\addtocounter{figure}{-1}
\begin{figure*}
  \resizebox{\hsize}{!}{\includegraphics[page=6,angle=0,trim=0.3cm 3cm 0.3cm 0.2cm,clip=true]{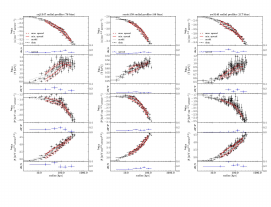}}
  \caption{continued.}
\end{figure*}

\end{appendix}

\end{document}